\documentclass[11pt,a4paper]{article}
\pdfoutput=1
\usepackage{jheppub}
\usepackage{amsmath}
\usepackage{dsfont}
\usepackage{hyperref}
\usepackage{ulem}
\usepackage{natbib}
\usepackage{xcolor}
\usepackage[hang,flushmargin]{footmisc}
\usepackage{tikz-cd}
\usepackage{enumitem}
\makeatletter\renewcommand{\@biblabel}[1]{#1.}\makeatother

\def\be{\begin{equation}}
\def\ee{\end{equation}}
\def\i{{\rm i}}
\def\e{{\rm e}}
\def\d{{\rm d}}

\newcommand{\ti}{{\text{i}}}
\DeclareMathOperator*{\Res}{Res}

\newcommand{\nn}{\nonumber}

\newcommand{\union}{{(\mathbb{C}_q \times \mathbb{S}^1) \cup (\mathbb{C}_{t^{-1}} \times \mathbb{S}^1)}}
\newcommand{\leftspace}{{\mathbb{C}_{t^{-1}} \times \mathbb{S}^1}}
\newcommand{\rightspace}{{\mathbb{C}_q \times \mathbb{S}^1}}

\newcommand{\mb}[1]{\mathbf{#1}}

\def\ket#1{{ |#1\rangle}}
\def\bra#1{{ \langle#1|}}
\def\braket #1#2{{ \langle #1|#2 \rangle}}

\preprint{
{\small{\textsf{UUITP-43/17}}}}

\title{3d Expansions of 5d Instanton Partition Functions}

\author{Fabrizio Nieri, Yiwen Pan and Maxim Zabzine}

\affiliation{Department of Physics and Astronomy, Uppsala University,\\
Box 516, SE-75120 Uppsala, Sweden.}

\emailAdd{fb.nieri@gmail.com}
\emailAdd{yiwen.pan@physics.uu.se}
\emailAdd{maxim.zabzine@physics.uu.se}

\abstract{We propose a set of novel expansions of Nekrasov's instanton partition functions. Focusing on 5d supersymmetric pure Yang-Mills theory with unitary gauge group on $\mathbb{C}^2_{q,t^{-1}} \times \mathbb{S}^1$, we show that the instanton partition function admits expansions in terms of partition functions of unitary gauge theories living on the 3d subspaces $\mathbb{C}_{q} \times \mathbb{S}^1$, $\mathbb{C}_{t^{-1}} \times \mathbb{S}^1$ and their intersection along $\mathbb{S}^1$. These new expansions are natural from the BPS/CFT viewpoint, as they can be matched with $\textrm{W}_{q,t}$ correlators involving an arbitrary number of screening charges of two kinds. Our constructions generalize and interpolate existing results in the literature. 
}


\begin{document}

\maketitle
\flushbottom

%%%%%% end of title page %%%%%%

%%%%%%%%%%%%%%%%%%%%%

\section{Introduction}

Since its debut \cite{Nekrasov:2002qd,Nekrasov:2003rj}, Nekrasov's instanton partition function, based on the works \cite{Losev:1997tp,Lossev:1997bz,Moore:1997dj,Moore:1998et}, has played a prominent role in subsequent development of supersymmetric gauge theories with $8$ supercharges in 4, 5 and 6 dimensions, as it concisely captures the non-perturbative physics of the gauge theories. As more studies are conducted, a handful of different representations are discovered in the contexts of supersymmetric gauge theories, topological vertex \cite{Aganagic:2003db,Iqbal:2007ii,Awata:2005fa}, two dimensional Liouville/Toda conformal field theories \cite{Alday:2009aq,Wyllard:2009hg}, and more. In this paper, we focus on the 5d $\Omega$-background and propose \textit{new expansions} in terms of codimension 2 and 4 partition functions, but most of our analysis can be extended to 4d and 6d as well.

The deep relations between 5d and 3d partition functions have been studied in a number of works, mainly in the context of codimension 2 BPS defects and the Higgsing procedure \cite{Gaiotto:2012xa,Gaiotto:2014ina,Nieri:2013vba,Pan:2016fbl,Gorsky:2017hro,Bullimore:2014awa} and large $N$ geometric transition or open/closed duality in refined topological strings \cite{Gopakumar:1998ki,Cachazo:2001jy,Aganagic:2002wv,Aganagic:2011mi}. At the practical level, the common denominator of the various approaches is that, upon appropriate \textit{limit} of the parameters, instanton partition functions reduce to vortex partition functions \cite{Hanany:2003hp,Shadchin:2006yz,Dimofte:2010tz,Bonelli:2011wx,Bonelli:2011fq,Aganagic:2013tta,Aganagic:2014oia,Aganagic:2014kja,Fujimori:2015zaa}. In this paper, we adopt a somewhat different perspective compared to the existing literature and observe a deeper connection between partition functions on $\mathbb{C}^2_{q,t^{-1}} \times \mathbb{S}^1$ and on $\mathbb{C}_q \times \mathbb{S}^1$ and/or $\mathbb{C}_{t^{-1}} \times \mathbb{S}^1$, even \textit{without} taking any limit.

\subsection{Summary of the results and motivations}
To give a brief summary of our results, we start by recalling one of the most frequently used representation of the instanton partition function of 5d $\mathcal{N} = 1$ $\textrm{U}(N)$ pure Yang-Mills theory on $\mathbb{C}^2_{q,t^{-1}} \times \mathbb{S}^1$, written as a sum over arbitrary Young diagrams $\vec Y=\{Y_A|A=1,\ldots,N\}$ labelling the fixed points of the instanton moduli space under the torus action ${\rm U}(1)_{\epsilon_1}\times {\rm U}(1)_{\epsilon_2}\times {\rm U}(1)^N_{\vec X}$. We have 
\be\label{ZY}
Z_\text{inst}(\vec x,Q_g;q,t)=\sum_{\vec Y}Z_\text{inst}^{\vec Y}(\vec x,Q_g;q,t) \ , 
\ee
where\footnote{We refer to \cite{Awata:2008ed} for more details and useful properties of Nekrasov's functions.}
\begin{align}
Z_\text{inst}^{\vec Y}(\vec x,Q_g;q,t)&=Q_g^{|\vec Y|}\prod_{A,B=1}^N\frac{1}{N_{Y_A Y_B}( x_{A}/x_{B};q,t)}~,\nn\\
N_{Y_A Y_B}( x;q,t)& = \prod_{(i,j)\in Y_A}(1-x q^{Y_{Ai}-j}t^{Y_{Bj}^\vee-i+1})\prod_{(i,j)\in Y_B}(1-x q^{-Y_{Bi}+j-1}t^{-Y_{Aj}^\vee+i})~,
\end{align}
and we have parametrized the Coulomb branch parameters with $x_{A}=\e^{2\pi\i X_A}$, the $\Omega$-background deformation parameters with $q=\e^{2\pi\i\epsilon_1}$, $t=\e^{-2\pi\i\epsilon_2}$ and the instanton counting parameter with $Q_g$. As usual, $Y_{Ai}$ denotes the length of the $i^{\rm th}$ row of $Y_A$, $|Y_A|$ denotes the number of boxes in  $Y_A$ with $|\vec Y| \equiv \sum_A |Y_A|$, while $Y_A^\vee$ denotes the transpose diagram.
%The array $\vec Y = \{Y_A|A = 1, \ldots, N\}$ denotes a collection of $N$ Young diagrams, with the length of the $i^{\rm th}$ row of $Y_A$ denoted by $Y_{Ai}$. $|Y_A|$ denotes the number of boxes in the diagram $Y_A$ and $|\vec Y| \equiv \sum_A |Y_A|$. These Young diagrams $\vec Y$ label the fixed points of the instanton moduli space under the torus action ${\rm U}(1)_{\epsilon_1}\times {\rm U}(1)_{\epsilon_2}\times {\rm U}(1)^N_{\vec x}$. 
%In this note, we will focus on the case of pure Yang-Mills as (anti-) fundamental or adjoint hypermultiplets can be easily added. as their contribution will be also described by the function $N_{YW}$, and their analysis will be similar to what we later discuss. Generalization to unitary quiver theories is also straightforward.
\begin{figure}[t]
  \centering
  \includegraphics[width=0.6\textwidth]{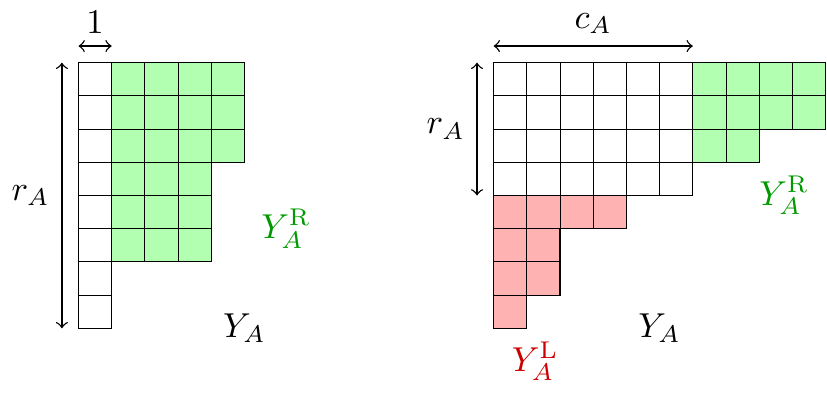}
  \caption{The first figure shows a particular diagram $Y_A \in \vec Y$ with exactly $r_A$ rows, which can be decomposed into the first column and a leftover $Y^\text{R}_A$ with at most $r_A$ rows. The second figure shows a diagram $Y_A$ containing a maximal rectangle (in white) of size $r_A \times c_A$ (such that $r_A - c_A = - 2$) in the upper-left corner, and two sub-diagrams $Y^\text{L}_A$ and $Y^\text{R}_A$ having at most $c_A$ columns and $r_A$ rows respectively. The transposed diagram $(Y^\text{L}_A)^\vee$ has at most $c_A$ rows. When $\vec{\mathfrak{n}} = \vec 0$, we return to the simpler case with maximal squares of shapes $r_A = c_A = d_A$.}\label{fig:Young-diagram-decomposition}
\end{figure}

We now observe that the instanton sum can be reorganized in several ways. An obvious organization, also frequently used, is as a sum over the \textit{instanton number} $k=|\vec Y|$, namely
\be\label{Zk}
Z_\text{inst}(\vec x,Q_g;q,t)=\sum_{k\geq 0} Z^k_\text{inst}(\vec x,Q_g;q,t) \ , \quad Z^k_\text{inst}(\vec x,Q_g;q,t)=\sum_{\substack{\vec Y\\ |\vec Y|=k}}Z_\text{inst}^{\vec Y}(\vec x,Q_g;q,t) \ .
\ee 
This is indeed the natural expansion arising from equivariant localization, and the summands can be nicely represented by a matrix model/contour integral computing the equivariant $\hat A$-genus on the instanton moduli space \cite{Losev:1997tp,Lossev:1997bz,Moore:1997dj,Moore:1998et}.  A less obvious expansion, which is our starting point,  organizes the instanton partition function as a sum over the \textit{number of rows} of the Young diagrams. If we denote by $\vec r=\{r_A | A = 1,\ldots,N\}$ the sequence of non-negative integers representing the number of {\it non-empty rows} in each diagram in $\vec Y$, we can write
\be\label{Zrows}
Z_\text{inst}(\vec x,Q_g;q,t)=\sum_{\vec r\in\mathbb{Z}^N_{\geq 0}} Z_\text{inst}^{\vec r}(\vec x,Q_g,t;q)~,
\ee
where $Z_\text{inst}^{\vec r}(\vec x,Q_g;q,t)$ captures all the contributions from the Young diagrams $\vec Y$ with {\it exactly} $\vec r$ rows (Figure \ref{fig:Young-diagram-decomposition}). As the notation suggests, this expansion breaks the $q \leftrightarrow t^{-1}$ symmetry explicitly. This symmetry can be restored by considering a yet another different expansion. In fact, for any Young diagram $Y_A \in \vec Y$ one can identify a maximal square in its upper-left corner of size $d_A \times d_A$ (Figure \ref{fig:Young-diagram-decomposition}). If we denote by $\mathbb{Y}[\vec d, \vec d]$ the set of Young diagrams $\vec Y$ having maximal squares of size $\{d_A \times d_A | A = 1,\ldots,N\}$, then clearly $\mathbb{Y}[\vec d, \vec d] \cap \mathbb{Y}[\vec d', \vec d'] = \emptyset$ whenever $\vec d \ne \vec d'$. Therefore, the sequence $\vec d$ characterizing the sizes of the maximal squares serves as a good organizing parameter, and we can organize the instanton sum as
\be
  Z_\text{inst}(\vec x,Q_g;q,t)=\sum_{\substack{\vec d\in\mathbb{Z}^N_{\geq 0}}}Z_\text{inst}^{\vec d,\vec d}(\vec x,Q_g;q,t)~.
\ee
We can readily generalize the above expansion by considering \textit{maximal rectangles} of shape $r_A \times c_A$ instead. We first fix a difference vector $\vec{\mathfrak{n}} \in \mathbb{Z}^N$. We denote by $\mathbb{Y}[\vec r, \vec c]$ the set of Young diagrams having their maximal rectangles of shape $\{r_A \times c_A|A=1\ldots N\}$ such that $\vec r - \vec c = \vec{\mathfrak{n}}$ (Figure \ref{fig:Young-diagram-decomposition}), which are frequently called hook diagrams. Clearly, $\mathbb{Y}[\vec r_1, \vec c_1] \cap \mathbb{Y}[\vec r_2, \vec c_2] = \emptyset$ if $\{\vec r_1, \vec c_2\} \ne \{\vec r_2, \vec c_2\}$ and $\vec r_1 - \vec c_1 = \vec r_2 - \vec c_2 = \vec{\mathfrak{n}}$. On the other hand, the union $\cup_{\vec r, \vec c | \vec r - \vec c = \vec {\mathfrak{n}}} \mathbb{Y}[\vec r, \vec c]$ exhausts all Young diagrams $\vec Y$. Therefore, we can also organize the instanton partition function for any fixed $\vec {\mathfrak{n}}$ as
\be\label{Zrowscol}
  Z_\text{inst}(\vec x,Q_g;q,t)= \sum_{\substack{(\vec r,\vec c)\in\mathbb{Z}^N_{\geq 0}\times \mathbb{Z}^N_{\geq 0}\\ \vec r - \vec c = \vec {\mathfrak{n}}}}Z_\text{inst}^{(\vec r,\vec c)}(\vec x,Q_g;q,t)~.
\ee
The main goal of this note is to sharpen the above observations and to study the physical and mathematical meaning of the different expansions. Our results include concrete expressions for the various summands, their gauge theory interpretation as partition functions of codimension 2 and 4 interacting theories on subspaces of $\mathbb{C}^2_{q,t^{-1}}\times\mathbb{S}^1$, and their BPS/CFT interpretation as the most general $\textrm{W}_{q,t}$ correlators. As we have mentioned, for the sake of clarity we will be mostly interested in pure Yang-Mills theory, but our analysis can be generalized to include matter and quiver theories.  

\subsection{Outline of the paper}
 In section \ref{subsection:new-expansion}, we study the concrete expression of Nekrasov's summands $1/\prod_{A,B}N_{Y_A Y_B}$ and show that they \textit{factorize} w.r.t. the decomposition of $\vec Y$ into left ($\vec Y^\text{L}$) and right ($\vec Y^\text{R}$) diagrams, see Figure \ref{fig:Young-diagram-decomposition} for an illustration.

In section \ref{subsection:identify-3d-partition-function}, we show that $Z_\text{inst}^{\vec r,\vec c}(\vec x,Q_g;q,t)$ admits a simple matrix model description, written as a contour integral (up to some explicit ``weight" factor)
\begin{align}
  Z_\text{inst}^{\vec r,\vec c}(\vec x,Q_g;q,t) \sim \oint \d^r z^\text{L} \d^c z^\text{R} Z^{\mathbb{C}_q \times \mathbb{S}^1}_{\textrm{U}(r), N}(z^\text{L}) Z^{\mathbb{S}^1}_\text{chiral}(z^\text{L}, z^\text{R}) Z^{\mathbb{C}_{t^{-1}} \times \mathbb{S}^1}_{\textrm{U}(c), N}(z^\text{R}) \,,
\end{align}
where $r=\sum_A r_A$, $c=\sum_A c_A$. This can be seen as generalized 3d holomorphic block integral \cite{Beem:2012mb}, where the  integrand includes the classical and 1-loop contributions from a pair of 3d $\mathcal{N}=2$ $\textrm{U}(r)$ and $\textrm{U}(c)$ gauge theories each coupled to one adjoint and $2N$ fundamental chiral multiplets on $\mathbb{C}_{q} \times \mathbb{S}^1$ and $\mathbb{C}_{t^{-1}} \times \mathbb{S}^1$ respectively, together with the 1-loop determinant of additional 1d $\mathcal{N} = 2$ chiral multiplets on $\mathbb{S}^1$ which transforms in the bifundamental representation of $\textrm{U}(r) \times \textrm{U}(c)$. The mass and FI parameters are also identified explicitly with the Coulomb branch and instanton parameters respectively.

\begin{figure}[t]
\centering
\includegraphics[width=0.8\textwidth]{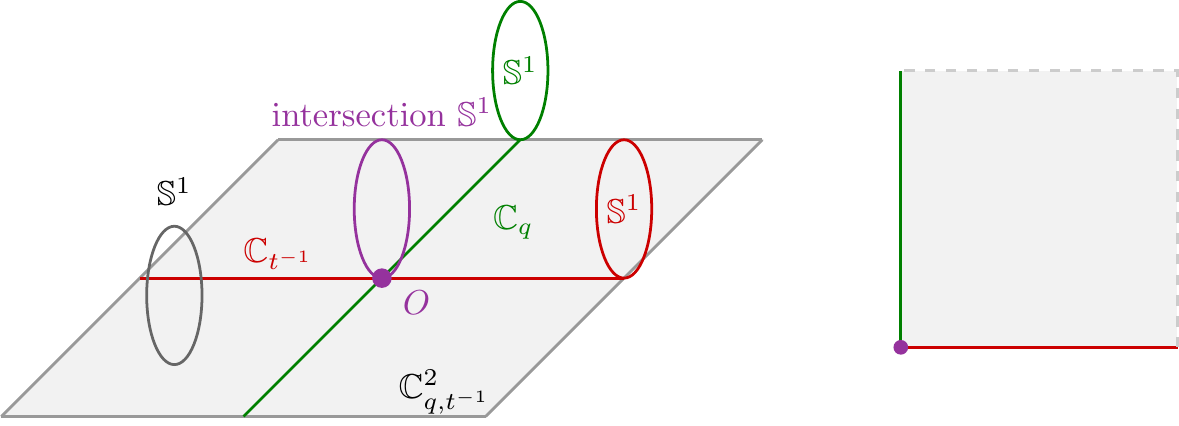}
\caption{A cartoon of $(\mathbb{C}_q \times \mathbb{S}^1 )\cup (\mathbb{C}_{t^{-1}} \times \mathbb{S}^1)$ as a subspace of $\mathbb{C}^2_{q,t^{-1}} \times \mathbb{S}^1$. We note that the two complex planes $\mathbb{C}_q$, $\mathbb{C}_{t^{-1}} \subset \mathbb{C}^2_{q,t^{-1}}$ actually intersect at the origin $O$. The bulk space $\mathbb{C}_{q,t^{-1}}^2 \times S^1$ can be represented by the toric diagram of the $\mathbb{T}^3$ action shown on the right. The $\mathbb{T}^3$ action reduces to $\mathbb{T}^2$ on the two edges corresponding to the subspaces $\mathbb{C}_q \times \mathbb{S}^1$ and $\mathbb{C}_{t^{-1}} \times S^1$, while it reduces to the rotation of the $\mathbb{S}^1$ at the vertex corresponding to the intersection $\mathbb{S}^1$.}\label{fig:cartoon}
\end{figure}

In section \ref{subsection:identify-3d-partition-function}, we argue that the above matrix model admits elegant interpretation as the partition function of a gauge theory living on the space $(\mathbb{C}_q \times \mathbb{S}^1) \cup (\mathbb{C}_{t^{-1}} \times \mathbb{S}^1)$ seen as a subspace of $\mathbb{C}^2_{q, t^{-1}} \times \mathbb{S}^1$. See Figure \ref{fig:cartoon}. Unlike the component spaces $\mathbb{C}_q \times S^1$ and $\mathbb{C}_{t^{-1}} \times S^1$, this space is not a smooth manifold. A gauge theory on such a space is given by three interacting ingredients: a 3d $\mathcal{N} = 2$ $\textrm{U}(r)$ gauge theory on $\mathbb{C}_q \times \mathbb{S}^1$, another similar $\textrm{U}(c)$ gauge theory on $\mathbb{C}_{t^{-1}} \times \mathbb{S}^1$, and an additional 1d $\mathcal{N} = 2$ theory living along the intersection $\mathbb{S}^1$. These three ingredients interact along the intersection $\mathbb{S}^1$ by coupling supersymmetrically via 1d $\mathcal{N} = 2$ superpotential and/or gauging, preserving the two supercharges of the 1d $\mathcal{N} = 2$ supersymmetry \cite{Gomis:2016ljm,Pan:2016fbl}. See also \cite{Nekrasov:2016qym,Nekrasov:2016ydq} for higher dimensional systems. 

In section \ref{sec:qVir}, we show that our new expansions are very natural from the viewpoint of the BPS/CFT correspondence \cite{Nekrasov:2012xe,Nekrasov:2013xda,Nekrasov:2015wsu}. In fact, we can match our results with a \textit{generating series} of $q$-Virasoro correlators involving an arbitrary number of screening charges of \textit{two kinds}. This correspondence generalize and interpolates between the constructions of \cite{Kimura:2015rgi} and \cite{Mironov:2011dk,Aganagic:2013tta}. In the former case, the $\mathbb{C}^2_{q,t^{-1}} \times \mathbb{S}^1$ instanton partition function is reproduced by considering an \textit{infinite} number of screening charges of only one kind. In the latter case, the $\mathbb{C}_{q} \times \mathbb{S}^1$ vortex partition function is reproduced by considering a \textit{finite} number of screening charges of only one kind, giving rise to the Dotsenko-Fateev matrix model representation, and the agreement between the approaches requires either fine tuning of the 5d Coulomb branch parameters or sending to infinity the rank of the 3d gauge group. 

The paper is supplemented with several appendixes where we collect useful definitions and technical computations.

\section{The three dimensional expansions}\label{sec:3dIPF}

\subsection{New expansions \label{subsection:new-expansion}}

As we recalled in the introduction, the instanton partition function of 5d $\mathcal{N} = 1$ $\textrm{U}(N)$ pure Yang-Mills theory on $\mathbb{C}^2_{q,t^{-1}} \times \mathbb{S}^1$ can be written as a sum over arbitrary Young diagrams
\begin{align}
  Z_\text{inst}(\vec x,Q_g;q,t)&=\sum_{\vec Y}Q_g^{|\vec Y|}\prod_{A,B=1}^N\frac{1}{N_{Y_A Y_B}(x_{AB};q,t)} \ ,
\end{align}
where we have used the shorthand notation $x_{AB}\equiv x_A/x_B$. The Nekrasov function $N_{Y_A Y_B}$ has a well-known representation in terms of $q$-Pochhammer symbols
\begin{align}
  N_{Y_A Y_B}(x;q,t)=\prod_{i,j=1}^\infty \frac{(x t^{j-i};q)_{Y_{Ai}-Y_{Bj}}}{(t\; x t^{j-i};q)_{Y_{Ai}-Y_{Bj}}}~.
\end{align}
If in $\vec Y$ each Young diagram $Y_A$ has \textit{at most} $r_A$ rows, the above product of $N_{Y_AY_B}$ can be written as
\begin{align}
  \prod_{A,B=1}^N\frac{1}{N_{Y_A Y_B}(x_{AB};q,t)}=\frac{\Delta_t(x_Y;q)}{\Delta_t( x_{\emptyset};q)}\prod_{B=1}^{N}\frac{V_t(x_Y,x_B t^{-r_B}; q)}{V_t( x_{\emptyset},x_B t^{-r_B};q)}~,
\end{align}
where the functions $\Delta_t(z;q)$ and $V_t(z, u;q,t)$ are defined in (\ref{Deltat}), (\ref{Vt}), with the collection of variables $x_Y$, $x_\emptyset$ given by
\begin{align}
  x_Y \equiv & \ \{x_A q^{Y_{Ai}}t^{1-i}~| A = 1, ..., N, \quad i = 1, ..., r_A\}~,\\
  x_\emptyset \equiv & \ \{x_A t^{1-i}~| A = 1, ..., N, \quad i = 1, ..., r_A\}~.
\end{align}
The upshot of this rewriting is that the resulting expression has the interpretation of the 1-loop determinant of a 3d $\mathcal{N}=2$ $\textrm{U}(r = \sum_A r_A)$ Yang-Mills theory coupled to one adjoint chiral multiplet with Neumann boundary conditions, $N$ fundamental chiral multiplets with Neumannt boundary conditions and $N$ fundamental chiral multiplets with Dirichlet boundary conditions, as one would derive from localization on $\mathbb{C}_q\times\mathbb{S}^1$ \cite{Yoshida:2014ssa}. Notice that the adjoint content is that of a 3d $\mathcal{N}=2^*$ theory. This motivates the definition of the partial sum over Young diagrams $\vec Y$ with all $Y_A$ having \textit{at most} $r_A$ rows, namely
\begin{align}\label{vortexlessr}
  Z_\text{inst}^{\le \vec r}(\vec x,Q_g,t;q)&=\sum_{\substack{\vec Y\\ \ell(\vec Y)\leq \vec r}}Q_g^{|\vec Y|}\frac{\Delta_t(x_Y;q)}{\Delta_t(\vec x_{\emptyset};q)}\prod_{B=1}^{N}\frac{V_t(x_Y,x_B t^{-r_B};q)}{V_t(\vec x_{\emptyset},x_B t^{-r_B};q)}~,
\end{align}
representing a vortex partition function for the theory we have just described, with the identification of the instanton counting parameter with the FI parameter. Then, the complete instanton partition function can be recovered by sending the rank of the 3d gauge group to infinity as
\begin{align}
  Z_\text{inst} (\vec x, Q_g; q,t) = \lim_{r_A \to +\infty} Z_\text{inst}^{\le \vec r}(\vec x,Q_g,t;q) \ .
\end{align}
Alternatively, we can define a closely related partial sum over only Young diagrams $\vec Y$ with each $Y_A$ having \textit{exactly} $r_A$ rows
\begin{align}
  Z_\text{inst}^{\vec r}(\vec x,Q_g,t;q)&=\sum_{\substack{\vec Y\\ \ell(\vec Y) = \vec r}}Q_g^{|\vec Y|}\frac{\Delta_t(x_Y;q)}{\Delta_t( x_{\emptyset};q)}\prod_{B=1}^{N}\frac{V_t(x_Y,x_B t^{-r_B};q)}{V_t( x_{\emptyset},x_B t^{-r_B};q)}~.
\end{align}
Then, the full instanton partition function can be recovered by summing over all $\vec r$
\begin{align}
  Z_\text{inst} (\vec x, Q_g; q,t) = \sum_{\vec r}Z_\text{inst}^{\vec r}(\vec x,Q_g,t;q) \ .
\end{align}
The above two approaches of reorganizing the instanton sum, though simple to implement, breaks the $q \leftrightarrow t^{-1}$ symmetry explicitly. In other words, the rows and columns are clearly not on the equal footing. From the geometry point of view, the original theory lives on $\mathbb{C}_q \times \mathbb{C}_{t^{-1}} \times \mathbb{S}^1$, while the above rewritings are related to vortex counting in three dimensional gauge theories living only on the submanifold $\mathbb{C}_q \times \mathbb{S}^1$.  

We thus task ourselves with finding some $q \leftrightarrow t^{-1}$ invariant expansions of the instanton partition function, in terms of 3d partition functions on both submanifolds $\mathbb{C}_q \times \mathbb{S}^1$ and $\mathbb{C}_{t^{-1}} \times \mathbb{S}^1$. It is crucial to point out that the two spaces actually intersect along a circle over the origin of both $\mathbb{C}_q$ and $\mathbb{C}_{t^{-1}}$. To implement this decomposition, we need to treat the rows and columns of the Young diagrams $\vec Y$ on equal footing. This suggests us to study the hook diagrams of type $(\vec r,\vec c)$ in more detail.

\begin{figure}[t]
  \centering
  \includegraphics[width=0.9\textwidth]{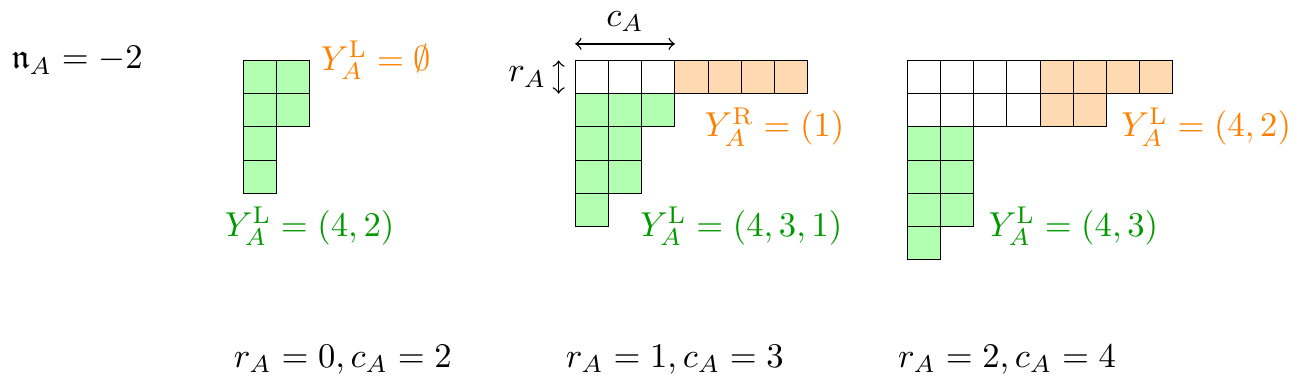}
  \caption{Examples of hook diagrams of various types with their maximal rectangles in white of shape $r_A \times c_A$, such that $\mathfrak{n}_A \equiv r_A - c_A = -2$. In the first example, the maximal rectangle is invisible due to the vanishing number of rows. The subdiagrams $Y^\text{L}_A$ and $Y^\text{R}_A$ are illustrated by colors.}\label{figure:maximal-rectangle}
\end{figure}

We begin by fixing a collection of integers $\vec {\mathfrak{n}} = \{\mathfrak{n}_A | A = 1, \ldots, N\}$. For any diagram $Y_A \in \vec Y$, we can always identify a unique maximal rectangle of shape $r_A \times c_A$ such that $r_A - c_A = \mathfrak{n}_A$, which simultaneously satisfies\footnote{\label{rcn}Note that these additional conditions are not always met by the maximal rectangle if the condition on $r_A, c_A$ is modified to $a r_A - b c_A = \mathfrak{n}_A$ for other integers $a, b \in \mathbb{Z}$.} 
\begin{align}
  Y_{Ai} \ge c_A, \quad i = 1, \ldots, r_A, \qquad \text{and} \qquad Y_{Ai} \le c_A, \quad i = r_A + 1, \ldots \ .\label{large-diagram}
\end{align}
Once the maximal rectangle is identified, we define the subdiagrams $\vec Y^\text{L}$ and $\vec Y^\text{R}$ of $Y$ by
\begin{align}
Y^\text{R}_{Ai} \equiv Y_{Ai} - c_A \ ,\quad i = 1, ..., r_A ~,\quad Y^\text{L}_{Ai} \equiv ~Y_{A(r_A + i)},\quad i = 1, ..., +\infty~.
\end{align}
Let us call the diagrams $\vec Y$ with the maximal rectangles $(\vec r,\vec c)$ hook diagrams of type $(\vec r, \vec c)$, the set of which denoted as $\mathbb{Y}[\vec r, \vec c]$. It is also convenient to rename $Y^\text{L} \to Y^{L\vee}$ such that the ``new" $Y^\text{L}_{A}$ has at most $c_A$ rows instead of columns. See Figure \ref{figure:maximal-rectangle} for simple examples, where the transposition $Y^\text{L}_{Ai}$ has been performed.
Clearly, $\mathbb{Y}[\vec r_1, \vec c_1] \cap \mathbb{Y}[\vec r_2, \vec c_2] = \emptyset$ if $\{\vec r_1, \vec c_2\} \ne \{\vec r_2, \vec c_2\}$ and $\vec r_1 - \vec c_1 = \vec r_2 - \vec c_2 = \vec{\mathfrak{n}}$, so that   the union $\cup_{\vec r, \vec c | \vec r - \vec c = \vec {\mathfrak{n}}} \mathbb{Y}[\vec r, \vec c]$ exhausts all Young diagrams $\vec Y$. We can now expand the instanton partition function as
\begin{align}
  Z_\text{inst}(\vec x,Q_g;q,t) &= \sum_{\substack{(\vec r, \vec c)\\ \vec r - \vec c = \vec {\mathfrak{n}}} }  Z^{\vec r, \vec c}_\text{inst}(\vec x,Q_g;q,t) \ , \nn\\
   Z^{\vec r, \vec c}_\text{inst}(\vec x,Q_g;q,t) & \equiv \sum_{\vec Y \in \mathbb{Y}[\vec r, \vec c]} Q_g^{|\vec Y|} \prod_{A,B} \frac{1}{N_{Y_A Y_B}(x_{AB};q,t)} \,,
\end{align}
and the only remaining problem is whether the product of Nekrasov functions behaves well under such new expansion. Without further ado, we claim that  (see appendix \ref{app:computations} for a derivation)\footnote{See also \cite{Pan:2016fbl} for similar factorization properties.}
\begin{multline} \label{main-claim}
 \frac{1}{\prod_{A,B}  N_{Y_A Y_B}(x_{AB};q,t)}=
   \frac{1}{N_{\square(\vec r, \vec c)}(\vec x)}\times\frac{V_{\rm int}( z_{ Y^\text{L}}, z_{ Y^\text{R}};p)}{V_{\rm int}( z_{ \emptyset^\text{L}}, z_{ \emptyset^\text{R}};p)}\times\\
   \times \Bigg[\frac{\Delta_t( z_{Y^\text{R}};q)}{\Delta_t( z_{\emptyset^\text{R}};q)}\prod_{B = 1}^N\frac{V_t( z_{Y^\text{R}},\eta^{\textrm{R}} x_B t^{-r_B} q^{c_B};q)}{V_t( z_{\emptyset^\text{R}},\eta^{\textrm{R}} x_B t^{-r_B} q^{c_B};q)} \Bigg]\Bigg[ (\text{R},\vec r,\vec c, q, t) \leftrightarrow (\text{L},\vec c,\vec r , t^{-1}, q^{-1})\Bigg] \ ,
\end{multline}
where we have defined:
\begin{itemize}[leftmargin=*]
\item the collections of variables
\begin{align}
  z_{Y^\text{R}} \equiv &\  \{z_{Y_{Ai}^\text{R}} = \eta^{\textrm{R}}x_A q^{c_A}q^{Y^\text{R}_{Ai}}t^{1-i}| A = 1, \ldots, N, \ i = 1, \ldots, r_A\} \ ,\label{right-poles}\\
  z_{Y^\text{L}} \equiv &\  \{z_{Y_{Ai}^\text{L}} = \eta^{\textrm{L}}x_A t^{ - r_A}t^{ - Y^\text{L}_{Ai}}q^{i - 1}| A = 1, \ldots, N, \ i = 1, \ldots, c_A\} \ ,\label{left-poles}
  \end{align}
and the parameters $\eta^{\textrm{R}},\eta^{\textrm{L}},p$ such that
\begin{align}
\eta^{\textrm{L}}/\eta^{\textrm{R}}\equiv(q t)^{1/2}\,,\quad p\equiv qt^{-1}\ ;
\end{align}

\item the intersection factor $V_\text{int}$ and the rectangle factor $N_{\square}$
\begin{align}
  V_{\rm int}( z_{ Y^\text{L}}, z_{ Y^\text{R}};p)& \equiv \prod_{A,B}\prod_{i=1}^{r_A}\prod_{j=1}^{c_B}\frac{1}{(1-p^{-1/2} z_{Y^\text{L}_{Bj}}/ z_{Y^\text{R}_{Ai}})(1-p^{-1/2} z_{Y^\text{R}_{Ai}}/z_{Y^\text{L}_{Bj}})} \ ,\\
  N_{\square(\vec r, \vec c)}(\vec x) & \equiv \ \prod_{A,B }\prod_{i=1}^{r_A}\frac{(x_{AB} t^{1-i};q)_{c_A-c_B}}{(x_{AB} t^{1+r_B-i};q)_{c_A-c_B}}\frac{(x_{AB} t^{r_B+1-i};q)_{c_A}}{(x_{AB}^{-1} t^{-r_B+i};q)_{-c_A}} \ .
\end{align}
\end{itemize}
The prefactor $N_{\square}$ captures the contribution from the maximal rectangle and, although it does not appear so, it is actually symmetric under $(\vec r,q,t)\leftrightarrow(\vec c, t^{-1}, q^{-1})$. In the next subsection, we give a matrix model description of this new expression which will help us to highlight its physical interpretation.

\subsection{The matrix model description\label{subsec:matrix-model}}

In the previous subsection, we have seen that the 5d  $\mathcal{N}=1$ $\textrm{U}(N)$ pure Yang-Mills theory can be expanded in a novel ways depending on a collection of integers $\vec {\mathfrak{n}}$ \footnote{From now on, when it is not necessary, the arguments of many functions will be omitted to avoid cluttering.}
\begin{align}
  Z_\text{inst} = \sum_{\substack{(\vec r, \vec c)\\ \vec r - \vec c = \vec {\mathfrak{n}}} } Z^{\vec r, \vec c}_\text{inst} \ .
\end{align}
More importantly, we have shown that the product of $N_{Y_A Y_B}$  factorizes neatly into ratios of functions $\Delta_t$,  $V_t$ (and their $q\leftrightarrow t^{-1}$ exchanged) which are very familiar in the context of vortex counting, along with some simple prefactor and intersection factor.

Two observations are in order. First of all, for fixed $\vec r, \vec c$, the above inner sum $\sum_{\vec Y \in \mathbb{Y}[\vec r, \vec c]}$ factorizes into a double sum, each of which is a sum over Young diagrams with at most $\vec r$ or $\vec c$ rows, namely
\begin{align}
  \sum_{\vec Y \in \mathbb{Y}[\vec r, \vec c]} = \sum_{\substack{\vec Y^\text{R} \\ \ell(\vec Y^\text{R}) \le \vec r}} \sum_{\substack{\vec Y^\text{L} \\ \ell(\vec Y^\text{L}) \le \vec c}} \ .
\end{align}
Second, the factorized combinations of $\Delta_t$, $V_t$ (and their $q\leftrightarrow t^{-1}$ exchanged) appearing in (\ref{main-claim}), together with the sums over Young diagrams with at most $\vec r$ ($\vec c$) rows, can be recast into an elegant matrix model.

Combining these two observations, we conclude that the contributions to the instanton partition function from all hook Young diagrams of type $(\vec r, \vec c)$ are captured by the matrix model
\begin{equation}\label{matrix-model}
  Z^{\vec r, \vec c}_\text{inst} \equiv \frac{Q_g^{\vec r\cdot \vec c}}{\mathcal{B}_{(\vec r, \vec c)}N_{\square(\vec r, \vec c)}} \oint_{\vec r, \vec c} \frac{\d^r z^\text{L}}{(2\pi \ti)^r} \frac{\d^c z^\text{R}}{(2\pi \ti)^c } \Upsilon_{q^{-1}}(z^\text{L};t^{-1}) V_\text{int}(z^\text{L}, z^\text{R}; p) \Upsilon_{t}(z^\text{R};q)\;,
\end{equation}
where the ranks are defined by $r \equiv \sum_{A = 1}^N r_A$, $c \equiv \sum_{A = 1}^N c_A$, $\vec r \cdot \vec c \equiv \sum_{A = 1}^N r_A c_A$, and:
\begin{itemize}[leftmargin=*]
\item  we have introduced two collections of variables 
\be 
z^\text{R} \equiv \{z^\text{R}_a | a = 1, \ldots, r\}\, , \quad z^\text{L} \equiv \{z^\text{L}_a | a = 1, \ldots, c\}\, ;
\ee
  
\item the $\Upsilon$ functions are defined as
\begin{align}
\Upsilon_t(z^\text{R};q)& \equiv  \Bigg(\prod_{a = 1}^r (z^\text{R}_a)^{\zeta^\text{R}-1} \Bigg) \Delta_t(z^\text{R}; q) \prod_{B = 1}^N V_t(z^\text{R}, \eta^{\textrm{R}} x_B t^{-r_B} q^{c_B};q)  ~\;,
\end{align}
where $\zeta^\text{R}$ and $\zeta^\text{L}$ are such that $q^{\zeta^\text{R}} = t^{- \zeta^\text{L}} = Q_g$, and the function $\Upsilon_{q^{-1}}(z^\text{L},;t^{-1})$ is defined similarly;

\item the intersection factor $V_\text{int}$ is defined as
\begin{align}
V_{\rm int}( z^\text{L}, z^\text{R};p)& \equiv \prod_{a=1}^r\prod_{b = 1}^{c}\frac{1}{(1-p^{-1/2} z_a^\text{R} /z^\text{L}_b)(1-p^{-1/2} z^\text{L}_b/z_a^\text{R})} \ ;\label{Vint-3d}
\end{align}

\item the integration contour is specified by selecting the poles given in (\ref{left-poles}) and (\ref{right-poles}).\footnote{These arise when integrating all the variables one after the other starting from the poles carried by the $V$ functions.} In particular, we recall that for $k \in \mathbb{Z}_{\geq 0}$, we have
\begin{align}
  \Res_{z = x q^{k}}\frac{1}{z(x/z;q)_\infty} = \frac{1}{(q^{-k};q)_{k}(q;q)_\infty}= (1;q)_{-k} \Res_{z=1}\frac{1}{z(z^{-1};q)_\infty} \ .
\end{align}
\end{itemize}
Finally, the coefficient $\mathcal{B}(\vec r, \vec c)$ is given by the residue
\begin{equation}
\mathcal{B}_{(\vec r, \vec c)}(\vec x,\zeta^{\textrm{L}},\zeta^{\textrm{R}}) = \Res_{\substack{z^\text{R} \to z_{\emptyset^\text{R}}\\ z^\text{L} \to z_{\emptyset^\text{L}}}} \Upsilon_{q^{-1}}(z^\text{L},\zeta^\text{L};t^{-1}) V_{\rm int}(z^\text{L},z^\text{R};p) \Upsilon_t(z^\text{R},\zeta^\text{R};q)  \ ,
\end{equation}
where $z_{\emptyset^\text{L,R}}$ is given by setting $Y^\text{L,R}$ to empty diagrams in $z_{Y^\text{L,R}}$. One can work out $1/\mathcal{B}N_{\square}$ explicitly, which reduces to
\begin{align}
  % \frac{Q^{\vec r\cdot \vec c}}{(\mathcal{B}N_{\square})(\vec r, \vec c)} = & \ Q^{\vec r\cdot \vec c} \ \Bigg[ \prod\limits_{A = 1}^N \prod\limits_{B = 1}^N \prod\limits_{i = 1}^{r_B} \frac{(x_{AB}t^{ + i};q)}{(tx_{AB}^{ - 1}t^{ - i};q)}\frac{(x_{AB}t^{i - {r_A}};q)}{(tx_{AB}^{ - 1}{t^{ - i + r_A}};q)}    \Bigg]\Bigg[ q\leftrightarrow t^{-1}, \vec r \leftrightarrow \vec c \Bigg] \nonumber\\
  \frac{1}{\mathcal{B}_{(\vec r, \vec c)}N_{\square(\vec r, \vec c)}} = &  \ \  \ \Bigg[ \prod_{A, B = 1}^N\prod_{i=0}^{r_A - r_B - 1} \frac{ (t \ x^{-1}_{AB} t^{+i} ;q)_\infty}{(x_{AB} t^{-i};q)_\infty} \Bigg]\Bigg[ q\leftrightarrow t^{-1}, \vec r \leftrightarrow \vec c \Bigg] \times\nonumber\\
  & \ \times \prod_{A, B = 1}^N \frac{
    \prod_{i=1}^{r_A - r_B}\prod_{j=1}^{c_A - c_B} (1- x_{AB} t^{1 - i}q^{j - 1})(1- x_{AB}^{-1} t^iq^{ - j})
  }{
    \prod_{i=0}^{r_B - r_A - 1}\prod_{j=0}^{c_A - c_B - 1} (1- x_{AB} t^{i + 1}q^j)(1- x_{AB}^{-1} t^{-i}q^{-j-1})
  } \times\nonumber\\
  & \ \times \Bigg[ \prod_{A = 1}^N \prod_{i=1}^{r_A} (\eta^{\textrm{R}} x_A t^{1-i} q^{c_A})^{-\zeta^\text{R}}\Bigg]\Bigg[\prod_{A = 1}^N \prod_{j=1}^{c_A} (\eta^{\textrm{L}} x_A q^{j-1} t^{-r_A})^{-\zeta^\text{L}}\Bigg]\times\nonumber\\
  & \ \times \Bigg[ \frac{1}{(t;q)_\infty} \Res_{z = 1} \frac{1}{z(z^{-1};q)_\infty} \Bigg]^r \Bigg[\frac{1}{(q^{-1};t^{-1})_\infty} \Res_{z = 1} \frac{1}{z(z^{-1};t^{-1})_\infty} \Bigg]^c \ .\label{prefactor}
\end{align}

In the next subsection, we will interpret our matrix model from the gauge perspective.

\subsection{Identification with \texorpdfstring{$(\mathbb{C}_q\times \mathbb{S}^1) \cup (\mathbb{C}_{t^{-1}}\times \mathbb{S}^1)$}{} partition functions \label{subsection:identify-3d-partition-function}}

\begin{figure}[t]
  \centering
  \includegraphics[width=0.6\textwidth]{intersection-defects-cartoon.pdf} ~~~ ~~~\includegraphics[width=0.3\textwidth]{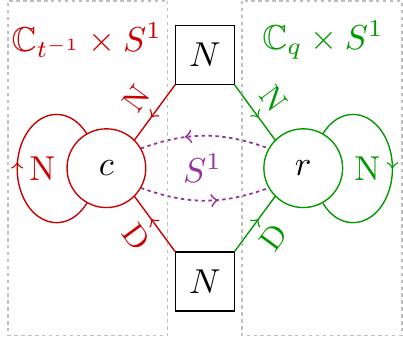}
  \caption{On the left is a cartoon and the toric diagram of the bulk space $\mathbb{C}_{q,t^{-1}}^2 \times \mathbb{S}^1$ and its subspace $(\mathbb{C}_{t^{-1}}\times \mathbb{S}^1) \cup (\mathbb{C}_q\times \mathbb{S}^1) \subset \mathbb{C}^2_{q,t^{-1}}  \times \mathbb{S}^1$, with the intersection given by $O \times \mathbb{S}^1 \subset \mathbb{C}^2_{q,t^{-1}} \times \mathbb{S}^1$. Both $\mathbb{C}_q\times \mathbb{S}^1$ and $\mathbb{C}_{t^{-1}}\times \mathbb{S}^1$ harbor respectively a 3d $\textrm{U}(r)$ and $\textrm{U}(c)$ gauge theory. The two gauge theories interact through a pair of 1d bifundamental chiral multiplets living at the intersection $\mathbb{S}^1$. On the right is the quiver diagram describing the intersecting gauge theories that enter into the expansion. The boundary conditions for various chiral multiplets are labeled explicitly, and the 1d chiral multiplets are denoted by the pair of purple dotted arrows in the middle.\label{fig:cartoon and quiver}}
\end{figure}

Now we are ready to interpret the matrix model (\ref{matrix-model}) in terms of 3d/1d gauge theory partition functions on the space $(\mathbb{C}_q \times \mathbb{S}^1) \cup (\mathbb{C}_{t^{-1}} \times \mathbb{S}^1) $ and to identify the physical parameters in these gauge theories. The union is specified as the setwise fixed points of the $\mathbb{T}^3$ action on $\mathbb{C}^2 \times \mathbb{S}^1$. We stress that it is a \textit{not} a smooth manifold, as the two components $\mathbb{C}_q \times \mathbb{S}^1$ and $\mathbb{C}_{t^{-1}} \times \mathbb{S}^1$, taken as two smooth submanifolds of $\mathbb{C}^2_{q, t^{-1}} \times \mathbb{S}^1$, actually \textit{intersect} along a circle $\mathbb{S}^1 = O \times \mathbb{S}^1$, where $O \in \mathbb{C}_q \cap \mathbb{C}_{t^{-1}}$ denotes the origin of $\mathbb{C}_{q,t^{-1}}$. See also the left of Figure \ref{fig:cartoon and quiver} for an illustration. 

As far as the individual component spaces  are concerned, partition functions of supersymmetric gauge theories on $\mathbb{D}^2_q \times \mathbb{S}^1\simeq \mathbb{C}_q \times \mathbb{S}^1$ can be studied by standard localization techniques \cite{Yoshida:2014ssa}. Such analysis presents the partition functions as the ``Coulomb branch" matrix models, a.k.a. 3d holomorphic block integrals \cite{Beem:2012mb,Pasquetti:2011fj}. It is straightforward to compare the integrand of the matrix model (\ref{matrix-model}) against the one-loop determinants in \cite{Yoshida:2014ssa}, which we collect in appendix \ref{app:index}. Indeed, the matrix model (\ref{matrix-model}) can be identified as
\be
  Z_\text{inst}^{\vec r, \vec c} =  \frac{Q_g^{\vec r\cdot \vec c}}{\mathcal{B}_{(\vec r, \vec c)} N_{\square(\vec r, \vec c)}}
 \oint_{\vec r, \vec c}\d^c \sigma^\text{L} \d^r\sigma^\text{R}Z^{\mathbb{C}_{t^{-1}} \times \mathbb{S}^1}_{\textrm{U}(c), N}(\sigma^\text{L}, m^\text{L}) Z^{\mathbb{S}^1}_\text{chiral}(\sigma^\text{L}, \sigma^\text{R}) Z^{\mathbb{C}_q \times \mathbb{S}^1}_{\textrm{U}(r), N} (\sigma^\text{R}, m^\text{R}) \ ,\label{partition-function-1}
  % = & \ \frac{W_r(m^\text{R};q)W_c(m^\text{L};t^{-1})}{(\mathcal{B} N_{\square})(\vec r, \vec c)} Z^{(\mathbb{C}_q\times \mathbb{S}^1) \cup (\mathbb{C}_{t^{-1}} \times \mathbb{S}^1)}_{\textrm{U}(r), \textrm{U}(c), N} \ .
\ee
where $Z^{\mathbb{C}_q \times \mathbb{S}^1}_{\textrm{U}(r), N}$ is the 1-loop determinant of the 3d $\mathcal{N}=2$ $\textrm{U}(r)$ gauge theory on $\mathbb{D}^2_q\times\mathbb{S}^1\simeq \mathbb{C}_q \times \mathbb{S}^1$ coupled to one Neumann adjoint (ad) chiral multiplet, $N$ Neumann (N) and  $N$ Dirichlet (D) fundamental chiral multiplets labeled by $A=1,\ldots, N$.\footnote{Alternatively, one can work with fundamental chirals satisfying the same boundary conditions but then ``boundary" interactions or Chern-Simons units are needed, see appendix \ref{app:index} and \cite{Yoshida:2014ssa} for more explanations.} Notice that the adjoint content is that of a 3d $\mathcal{N}=2^*$ vector multiplet. Similarly for $Z^{\mathbb{C}_{t^{-1}} \times \mathbb{S}^1}_{\textrm{U}(c), N}$, while 
\begin{align}
  Z_\text{chiral}^{\mathbb{S}^1}(\sigma^\text{L}, \sigma^\text{R})  \equiv \prod_{a=1}^r\prod_{b = 1}^c \prod_\pm\frac{p^{1/2} }{2\sinh \pi \i\big(\pm(\sigma_b^\text{L} - \sigma_a^\text{R}) + \frac{1}{2}(\epsilon_1 + \epsilon_2)\big)}
\end{align}
is the 1-loop determinant of a pair of \textit{native} 1d $\mathcal{N} = 2$ chiral multiplets living on the intersection circle $O \times \mathbb{S}^1$ and transforming in the bifundamental representation of the gauge group $\textrm{U}(r) \times \textrm{U}(c)$. Here, we have identified $z^{\textrm{L,R}}=\e^{2\pi\i\sigma^{\textrm{L,R}}}$.  Introducing the parametrization $\eta^{\textrm{L,R}} \equiv \e^{2\pi \i \hat \eta^{\textrm{L,R}}}$,  and recalling the definitions $q\equiv \e^{2 \pi \i \epsilon_1}$, $t \equiv \e^{ - 2 \pi \i \epsilon_2}$, $x_A \equiv \e^{2\pi \i X_A}$, the mass parameters ($m$) of the 3d theories are 
\begin{align}
  & m^\text{R,N}_A = X_A + c_A \epsilon_1 + (r_A - 1)\epsilon_2  +\hat\eta^\textrm{R}\ , \ \ m^\text{R,D}_A = X_A + (c_A + 1) \epsilon_1 + r_A \epsilon_2 +\hat\eta^\textrm{R} \ , \label{mass-id-1}\\
  & m^\text{L,N}_A = X_A + r_A \epsilon_2 + (c_A - 1)\epsilon_1 +\hat\eta^\textrm{L}\ , \ \ m^\text{L,D}_A = X_A + (r_A + 1) \epsilon_2 + c_A \epsilon_1 +\hat\eta^\textrm{L} \ \label{mass-id-2}\ ,\\
  & m_{\textrm{ad}}^\text{R} = - \epsilon_2, \ \ m_{\textrm{ad}}^\text{L} = - \epsilon_1 \ ,
\end{align}
and both theories have non-trivial FI parameters given by
\begin{align}
  \xi_\text{FI}^\text{R} &=  \ \zeta ^\text{R} \ ,\qquad \xi_\text{FI}^\text{L} =  \ \zeta ^\text{L} \ .
\end{align}
% \begin{align}
%   \xi_\text{FI}^\text{R} &=  \ \zeta ^\text{R}  - \frac{1}{2\epsilon _1}\sum_{A = 1}^N (  \epsilon_1 + m_A^\text{R,N} - m_A^\text{R,D}) \ ,\quad \xi_\text{FI}^\text{L} =  \ \zeta ^\text{L}  - \frac{1}{2\epsilon _2}\sum_{A = 1}^N (  \epsilon_2 + m_A^\text{L,N} - m_A^\text{L,D}) \ .
% \end{align}
The analysis of the normalization of the matrix model is rather involved and we refer interested readers to appendix \ref{app:freesec}. Essentially, it corresponds to a free sector.

In the beginning of this subsection, we have anticipated that the matrix integral (\ref{partition-function-1}) admits an interpretation as the partition function of certain 3d gauge theory on the space $(\mathbb{C}_q \times \mathbb{S}^1) \cup (\mathbb{C}_{t^{-1}} \times \mathbb{S}^1) $. Defining supersymmetric gauge theories on intersecting spaces, $\union$ in our example, is straightforward and was explored in great detail in \cite{Nekrasov:2016qym,Nekrasov:2016ydq,Gomis:2016ljm,Pan:2016fbl}. Here we summarize relevant aspects. On both $\leftspace$ and $\rightspace$, we define respectively 3d $\mathcal{N} = 2$ $\textrm{U}(c)$ and $\textrm{U}(r)$ gauge theories referred to as $\mathcal{T}^\text{L}$ and $\mathcal{T}^\text{R}$ in the usual manner: away from the intersection $\mathbb{S}^1$, both quantum field theories separately behave just normally. The two gauge theories should, however, interact along the intersection $\mathbb{S}^1$. To capture this interaction, we place an additional 1d $\mathcal{N} = 2$ theory $\mathcal{T}^\text{1d}$ of a collection of 1d $\mathcal{N} = 2$ supermultiplets. Along the $\mathbb{S}^1$, we decompose all the 3d $\mathcal{N} = 2$ supermultiplets in both $\mathcal{T}^\text{L}$ and $\mathcal{T}^\text{R}$ in terms of 1d $\mathcal{N} = 2$ supermultiplets. In particular, we have the pattern of decomposition summarized in the following table:
\begin{center}
  \begin{tabular}{c|c}
    3d $\mathcal{N} = 2$ multiplet $\Phi^\text{L,R}$ & 1d $\mathcal{N} = 2$ multiplets $\phi^\text{L,R}$ after decomposition\\
    \hline
    vector       & vector and Fermi \\
    chiral       & chiral and Fermi \\
  \end{tabular} \ .
\end{center}
Once the supermultiplets in $\mathcal{T}^\text{L,R}$ are decomposed along the intersection $\mathbb{S}^1$, the resulting 1d $\mathcal{N} = 2$ components can couple to the supermultiplets in $\mathcal{T}^\text{1d}$ in supersymmetric fashion preserving the 1d $\mathcal{N} = 2$ supersymmetry on $\mathbb{S}^1$: the 1d $\mathcal{N} = 2$ vector multiplets from $\mathcal{T}^\text{L,R}$ can gauge the global symmetry of $\mathcal{T}^\text{1d}$, while the 1d $\mathcal{N} = 2$ chiral and Fermi multiplets from $\mathcal{T}^\text{L,R}$ can couple to those in $\mathcal{T}^\text{1d}$ via superpotentials $W^\text{3d/1d}$. Note that, although being $\mathcal{Q}$-exact and therefore do not actually enter into the localization computation, superpotentials will impose relations between masses and $\textrm{U}(1)_\mathcal{R}$ charges across theories in different dimensions. The final product is then an action $S^\text{3d/1d}$ describing the 3d/1d coupled system
\begin{multline}
  S^\text{3d/1d}[\Phi^\text{L,R},\phi^\text{1d}] = S^{\mathbb{C}_{t^{-1}} \times \mathbb{S}^1}_{\mathcal{T}^\text{L}}[\Phi^\text{L}] + S^{\mathbb{C}_q \times \mathbb{S}^1}_{\mathcal{T}^\text{R}}[\Phi^\text{R}] +\\
  + S^{\mathbb{S}^1}_{\mathcal{T}^\text{1d}}[\text{vm}^\text{L,R}, \phi^\text{1d}] +\int_{\mathbb{S}^1} W^\text{3d/1d}(\Phi^\text{L,R}|_{\mathbb{S}^1}, \phi^\text{1d}) \ . 
\end{multline}
Here, we have explicitly introduced 1d vector multiplets vm$^\text{L,R}$ from $\mathcal{T}^\text{L,R}$ to indicate the gauging of the global symmetry of $\mathcal{T}^\text{1d}$. The partition function of the overall gauge theory on $\union$ is defined  by the path integral
\begin{align}
  Z^\union = \int \Big[D\Phi^\text{L,R} \Big] \Big[D\phi^\text{1d}\Big]\, \e^{ - S^\text{3d/1d} [\Phi^\text{L,R}, \phi^\text{1d}]} \ .
\end{align}
Supersymmetric localization can be performed by first localizing the 1d theory, then the 3d theories, allowing one to use standard techniques in this setup too.

From the matrix model (\ref{partition-function-1}), we can recognize $\mathcal{T}^\text{R}$ to be the $\textrm{U}(r)$ gauge theory coupled to the aforementioned collection of chiral multiplets, together with a collection of free chiral multiplets, and similarly for $\mathcal{T}^\text{L}$. On the intersection $\mathbb{S}^1$, the 1d $\mathcal{N} = 2 $ theory $\mathcal{T}^\text{1d}$ consists of a pair of chiral multiplets transforming in the bifundamental representation of $\textrm{U}(r) \times \textrm{U}(c)$\footnote{In other words, a $\textrm{U}(r) \times \textrm{U}(c)$ subgroup of the global symmetry of $r \times c$ free 1d $\mathcal{N} = 2$ chiral multiplets is gauged by the vector multiplets in $\mathcal{T}^\text{L,R}$.}, together with a collection of free 1d chiral and Fermi multiplets. See the right of Figure \ref{fig:cartoon and quiver} for the quiver structure of the interacting sector. From (\ref{mass-id-1}) and (\ref{mass-id-2}), we notice $ \epsilon_1 + m^\text{R,N} - m^\text{R,D} = - \epsilon_2 $ and $  \epsilon_2 + m^\text{L,N} - m^\text{L,D} = - \epsilon_1$. We also recall $\eta^{\textrm{L}}/\eta^{\textrm{R}}  = (qt)^{1/2}$. We are then immediately lead to the left/right mass relations,
\begin{align}
  m^\text{R,N}_A - m^\text{L,N}_A = + \frac{1}{2}(\epsilon_1 - \epsilon_2), \quad m^\text{R,D}_A - m^\text{L,D}_A = + \frac{1}{2}(\epsilon_1 - \epsilon_2), \quad m^\text{R}_{\textrm{ad}} - m^\text{L}_{\textrm{ad}} = \epsilon_1 - \epsilon_2 \ .
\end{align}
Note that the masses denoted by $m$ are the complex combinations of the real masses and the $\textrm{U}(1)_\mathcal{R}$ charges. We are thus naturally led to combine the matrix integral and the free theory contributions inside $1/\mathcal{B}N_{\square}$, and denote the whole object as $Z^\union_{\textrm{U}(r), \textrm{U}(c), N}$.\footnote{When evaluating the integral, the contour depends on $\vec r, \vec c$.} Finally, the instanton partition function of 5d $\mathcal{N}=1$ $\textrm{U}(N)$ pure Yang-Mills theory can be expanded in terms of $Z^\union_{\textrm{U}(r), \textrm{U}(c), N}$ as
\begin{align}
  Z_\text{inst} = \sum_{\substack{(\vec r, \vec c)\\ \vec r - \vec c = \vec {\mathfrak{n}}} } W_{\vec r,\vec c} \, Z^\union_{\textrm{U}(r), \textrm{U}(c), N} \ ,
  \label{final-expansion}
\end{align}
where $W_{r,c}$ is a (sufficiently simple) ``weight" factor given in appendix \ref{app:freesec}.

\textbf{Remark}. The expansion (\ref{Zrows}), where one sums only over the rows of the Young diagrams, corresponds to the particular (degenerate) case where one fixes $c_A=1$ and picks up only the poles labeled by  $z_{\emptyset^\textrm{L}}$. In the notation of footnote \ref{rcn}, this corresponds to $a=0,b=1,\mathfrak{n}_A=-1$. In this case, the dynamics on the $\mathbb{C}_{t^{-1}}\times\mathbb{S}^1$ subspace is trivial, and the $\mathbb{C}^2_{q,t^{-1}}\times\mathbb{S}^1$ instanton partition function can entirely be described by the $\textrm{U}(r)$ theory on $\mathbb{C}_{q}\times\mathbb{S}^1$. With no interactions between the two orthogonal subspaces, also the free sector is much simpler, and in the prefactor (\ref{prefactor}) only the terms with $c_A=1$ survive, with the second line disappearing completely.

% Finally, the FI-parameters are also related by $\epsilon_1 \xi_\text{FI}^\text{R} = \epsilon_2 \xi_\text{FI}^\text{L}$, thanks to the definitions $Q = q^{\zeta^\text{R}} - t^{- \zeta^\text{L}} = \frac{1}{2}(\epsilon_1 - \epsilon_2)$ which implies $\epsilon_1 \zeta^\text{R} = \epsilon_2 \zeta^\text{L}$. The relation between the two FI-parameters does not follow from the gauge theory construction, however, it could be given geometric interpretation if one realizes the above gauge theory on $\union$ by suspending $r$ and $c$ D3 branes between a pair of parallel NS5 branes.

\section{\texorpdfstring{$q$}-Virasoro correlators}\label{sec:qVir}
In this section, we show that our new expansions are natural from the viewpoint of the BPS/CFT correspondence too. As a byproduct, we will establish a precise connection between two slightly different approaches in existing literature. This observation is closely related to \cite{Carlsson:2013jka}. On the one hand, the $\mathbb{C}^2_{q,t^{-1}} \times \mathbb{S}^1$ instanton partition function of 5d $\mathcal{N}=1$ $\textrm{U}(N)$ Yang-Mills theory (possibly coupled to (anti-) fundamental matter) can be given as a free boson correlator involving \textit{infinitely-many} screening charges $\mb{Q}^{(+)}$ (possibly together with vertex operators) of the $q$-Virasoro algebra \cite{Kimura:2015rgi}. On the other hand, the $\mathbb{C}_{q} \times \mathbb{S}^1$ vortex partition function of the 3d $\mathcal{N}=2$ $\textrm{U}(r)$ Yang-Mills theory coupled to one adjoint chiral (and possibly to (anti-) fundamental chiral matter) can be given as a free boson correlator involving \textit{finitely-many} $r$ screening charges $\mb{Q}^{(+)}$, possibly with vertex operators \cite{Aganagic:2013tta}. It is known that, in the presence of enough amount of fundamental hyper multiplets, the two descriptions agree upon taking the 5d equivariant parameter $x_A $ to special values $ x_A^*$ which depends on the hyper multiplet masses. Such a limit is closely related to Higgsing as described in \cite{Gaiotto:2012xa}. This is usually seen as an equivalence, in the sense that, when the setup is embedded in String/M-theory, one can safely switch from one phase to the other by large $r$ open/closed string duality or geometric transition. Below, we are going to show that a similar relation continues to hold without taking any specialization/limit and even when the 5d theory cannot be Higgsed, and  simultaneously preserve the $q\leftrightarrow t^{-1}$ symmetry which would have been broken by a choice of  a preferred $\mathbb{C}$ plane in $\mathbb{C}^2 \times \mathbb{S}^1$. For the sake of completeness and to fix our conventions, we first briefly review the free boson representation of the $q$-Virasoro algebra and then compute correlators with finitely-many screening charges. The comparison with the (less standard) approach involving infinitely-many screening charges is presented in appendix \ref{KPapp}.

\subsection{Screening currents and vertex operators}
Consider the Heisenberg algebra generated by oscillators $\{\mb{a}_m, m \in \mathbb{Z}\backslash \{0\} \}$ and zero modes $\mb{P}, \mb{Q}$, with the non-trivial commutation relations 
\begin{align}
 \big[\mb{a}_m, \mb{a}_n\big] &= -\frac{1}{m}(q^{m/2} - q^{-m/2})(t^{-m/2} - t^{m/2})C^{[m]}(p)\delta_{m+n, 0}\;, \qquad \big[\mb{P}, \mb{Q} \big] = 2\ ,
\end{align}
where $C^{[m]}(p)=(p^{m/2} + p^{-m/2})$ is the deformed Cartan matrix of the $A_1$ algebra. Here, $q,t\in\mathbb{C}$ and $p=q t^{-1}$. The $q$-Virasoro current $\mb{T}(z)=\sum_{m\in\mathbb{Z}}\mb{T}_m z^{-m}$ can be realized as
\be
  \mb{T}(z) = \mb{Y}(p^{-1/2}z) + \mb{Y}(p^{1/2}z)^{-1}, \quad \mb{Y}(z) = \ : \exp \Bigg[ \sum_{m \ne 0} \frac{\mb{a}_m  \ z^{-m}}{C^{[m]}(p)}\Bigg]q^{\sqrt{\beta}\mb{P}/2}p^{1/2} :\ , 
  \label{stress-tensor-freeboson}
\ee
where $\beta\in\mathbb{C}$ is such that $t\equiv q^\beta$ and the normal ordering $: ~ :$ pushes the positive oscillators and $\mb{P}$ to the right. The screening currents of the $q$-Virasoro algebra have the following free boson representation 
\begin{align}\label{standard-screening-current}
  \mb{S}^{(\pm)}(z)& \equiv  \ :\exp\Bigg[ - \sum_{m \ne 0} \frac{\mb{a}_m \ z^{-m}}{ \mathfrak{q}_\pm^{m/2} - \mathfrak{q}_\pm^{- m/2}} \pm\sqrt{\beta^{\pm 1}}\mb{Q}\pm\sqrt{\beta^{\pm 1}}\mb{P}\ln z\Bigg]: \ ,
\end{align}
where $\mathfrak{q}_+\equiv q$, $\mathfrak{q}_-\equiv t^{-1}$. Their defining property is
\be\label{Scurr}
\Big[\mb{T}_m,\mb{S}^{(\pm)}(z)\Big]=\frac{\widehat{T}_{\mathfrak{q}_\pm}-1}{z}\mathcal{O}_m(z) \ ,
\ee
where we have defined a shift operator acting as $\widehat{T}_{\mathfrak{q}_\pm} f(z)=f(\mathfrak{q}_\pm z)$. For a given $\gamma\in \mathbb{C}$ and $u\equiv q^{\sqrt{\beta}\gamma}$, we define the vertex operators 
\begin{align}
\mb{V}(x)&\equiv \ : \exp\Bigg[-\sum_{m\neq 0}\frac{1 }{(q^{m/2}-q^{-m/2})(t^{m/2}-t^{-m/2})}\frac{\mb{a}_{m}\ x^{-m}}{C^{[m]}(p)}\Bigg]: \ ,\\
\mb{H}_u(x)&\equiv \ :\exp\Bigg[-\sum_{m\neq 0}\frac{(u^{-m}-u^{m})}{(q^{m/2}-q^{-m/2})(t^{m/2}-t^{-m/2})}\frac{\mb{a}_{m}\ x^{-m}}{C^{[m]}(p)}+\frac{\gamma}{2}\mb{Q}+\frac{\gamma}{2}\mb{P}\ln x\Bigg]: \ .
\end{align}
The interesting ``OPE"  of screening currents and vertex operators are as follows 
\begin{align}
\mb{S}^{(\pm)}(z)\mb{S}^{(\pm)}(w)&=\ :\mb{S}^{(+)}(z)\mb{S}^{(+)}(w): \Delta_{\mathfrak{q}^{-1}_\mp} (z,w;\mathfrak{q}_\pm) \ (zw)^{\beta^{\pm 1}} \ c_{\beta^{\pm 1}}(z,w;\mathfrak{q}_\pm)\ , \\
\mb{S}^{(-)}(z)\mb{S}^{(+)}(w)&= \ :\mb{S}^{(-)}(z)\mb{S}^{(+)}(w): \ \frac{(-p^{1/2}zw)^{-1}}{(1-p^{-1/2}z/w)(1-p^{-1/2}w/z)}\ ,\\
\mb{S}^{(\pm)}(z)\mb{V}(x)&=\ :\mb{S}^{(\pm)}(z)\mb{V}(x): \ ( \mathfrak{q}_\pm^{1/2} x/z ;\mathfrak{q}_\pm)_\infty  \ ,\\
\mb{V}(x)\mb{S}^{(\pm)}(z)&=\ :\mb{S}^{(\pm)}(z)\mb{V}(x): \ \frac{1}{(\mathfrak{q}_\pm^{1/2} z/x ;\mathfrak{q}_\pm)_\infty} \ , \\
\mb{H}_u(x)\mb{S}^{(\pm)}(z)&=\ :\mb{H}_u(x)\mb{S}^{(\pm)}(z): \ \frac{(\mathfrak{q}^{1/2}_\pm z u/ x  ;\mathfrak{q}_\pm)_\infty}{(\mathfrak{q}^{1/2}_\pm z/ x u ;\mathfrak{q}_\pm)_\infty} \ x^{\pm\gamma\sqrt{\beta^{\pm 1}}} \ ,
\end{align}
where we defined the functions 
\begin{align}
\Delta_{\mathfrak{q}^{-1}_\mp} (z,w;\mathfrak{q}_\pm)&\equiv\frac{(z/w;\mathfrak{q}_\pm)_\infty (w/z;\mathfrak{q}_\pm)_\infty}{(\mathfrak{q}^{-1}_\mp z/w;\mathfrak{q}_\pm)_\infty(\mathfrak{q}^{-1}_\mp w/z;\mathfrak{q}_\pm)_\infty}\ ,\nn\\
c_{\beta^{\pm 1}}(z,w;\mathfrak{q}_\pm)&\equiv\frac{\Theta( \mathfrak{q}^{\beta^{\pm 1}}_\pm z/w;\mathfrak{q}_\pm)}{\Theta(z/w;\mathfrak{q}_\pm)}\ \left(\frac{z}{w}\right)^{\beta^{\pm 1}} \ .
\end{align}
Finally, for any given $\alpha\in\mathbb{C}$, we consider the left and right Fock modules over the charged Fock vacua $\ket{\alpha}=\e^{\alpha\mb{Q}/2}\ket{0}$ and $\bra{\alpha}=\bra{0}\e^{-\alpha\mb{Q}/2}$ respectively, namely
\be
\mb{P}\ket{\alpha}=\alpha\ket{\alpha} \ , \quad \mb{a}_m\ket{\alpha}=0\ ,\quad \bra{\alpha}\mb{a}_{-m}=0 \ ,\quad m\in\mathbb{Z}_{>0}\ ,
\ee 
with $\braket{0}{0}=1$. We are now ready to compute various $q$-Virasoro collelators.

\subsection{Finitely-many screening currents}

Recall that the commutator between $\mb{T}_m$ and $\mb{S}(z)$ is a total difference $z^{-1}(\mathcal{O}(qz) - \mathcal{O}(z))$ for some fixed operator $\mathcal{O}(z)$. Therefore, for contours\footnote{For instance, one can take the contour to circle the poles in the meromorphic factors arising from normal-ordering the product of $\mb{S}$.} of $z_i$ invariant under $q$-shifts, the integrated product of screening currents
\begin{align}
  \Big[\mb{Q}^{(\pm)}\Big]^{r} \equiv \oint dz_1\ldots dz_r \mb{S}^{(\pm)}(z_1)\ldots\mb{S}^{(\pm)}(z_r)
\end{align}
will be annihilated by $\mb{T}_m$ in commutator, thanks to $dz/z = d(qz)/(qz)$.

Let us now consider this operator and perform the normal ordering for the screening currents,
\be
\mb{Z}_r\equiv \Big[\mb{Q}^{(+)}\Big]^{r}=\oint\Big[\prod_{i=1}^{r}\frac{\d z_i}{2\pi\i z_i}\, z_i^{\sqrt{\beta}(\sqrt{\beta}r-Q)}\Big]\, c_\beta(z;q)\,\Delta_t(z;q)\,:\prod_{i=1}^r \mb{S}^{(+)}(z_i):\ ,
\ee
where $Q\equiv \sqrt{\beta}-1/\sqrt{\beta}$. Notice that we have explicitly broken the $q\leftrightarrow t^{-1}$ symmetry by considering only one kind of screening charge, and we have considered finitely-many insertions in order to have a conventional finite rank matrix model, with potential parametrized by the coefficients $\{\mb{a}_{-m},m >0\}$. We can now compute the normalized correlator
\begin{multline}\label{HZvortex}
\frac{\bra{\alpha_\infty}\prod_{f=1}^{N_\textrm{f}}\mb{H}_{u_f}(x_f)\mb{Z}_r\ket{\alpha_0}}{\bra{\sum_f\gamma_f}\prod_{f=1}^{N_\textrm{f}}\mb{H}_{u_f}(x_f)\ket{0}} =\\
=\oint\Big[\prod_{i=1}^{r}\frac{\d z_i}{2\pi\i z_i}\, z_i^{\sqrt{\beta}(\alpha_0+\sqrt{\beta}r-Q)}\Big]\, c_\beta(z;q)\, \Delta_t(z;q)\, \prod_{i,f}\frac{(q^{1/2}u_f z_i/x_f ;q)_\infty}{(q^{1/2} z_i /u_fx_f ;q)_\infty} \ ,
\end{multline}
where $\alpha_\infty=\alpha_0+2\sqrt{\beta}r+\sum_f\gamma_f$ for charge conservation. This has the form of a Dotsenko-Fateev matrix model. As follows from the BPS/CFT correspondence, in the expression above we can easily recognize the block integral for the vortex partition function of the 3d $\mathcal{N}=2$ $\textrm{U}(r)$ Yang-Mills theory coupled to one adjoint and $N_\textrm{f}$ fundamental and anti-fundamental chirals, with FI parameter $\sqrt{\beta}(\alpha_0+\sqrt{\beta}r-Q)$ \cite{Aganagic:2013tta}.\footnote{\label{cbeta}One should observe that the $c_\beta$ function reduces to an overall constant on the chosen integration contour.} This matrix model also corresponds to the Nekrasov instanton partition function of 5d $\mathcal{N}=1$ $\textrm{U}(N)$ Yang-Mills theory coupled to $N$ fundamental and $N$ anti-fundamental matter at specific points in the Coulomb branch (see appendix \ref{KPapp}).

\subsection{Generating series of correlators}
In this subsection, we generalize the above computation to include both types of screening charges, and we establish the correspondence with the new Nekrasov expansions studied in the previous section. Let us start by considering the most general operator constructed with a finite number of $q$-Virasoro screening charges
\begin{multline}
\mb{Z}_{(r_-,r_+)}\equiv \Big[\mb{Q}^{(-)}\Big]^{r_-}\Big[\mb{Q}^{(+)}\Big]^{r_+}=\oint\prod_\pm\prod_{i=1}^{r_\pm}\frac{\d z_{\pm,i}}{2\pi\i z_{\pm ,i}}\, z_{\pm ,i}^{\pm\sqrt{\beta^{\pm 1}}\alpha_{(r_-,r_+)}}\times\\
\times \frac{\prod_\pm c_{\beta^{\pm 1}}(z_\pm;\mathfrak{q}_\pm)\,\Delta_{\mathfrak{q}_\mp^{-1}}(z_\pm;\mathfrak{q}_\pm)}{\prod_{i=1}^{r_+}\prod_{j=1}^{r_-}(-p^{1/2})(1-p^{-1/2}z_{-,j}/z_{+,i})(1-p^{-1/2}z_{+,i}/z_{-,j})}\prod_\pm:\prod_{i=1}^{r_\pm} \mb{S}^{(\pm)}(z_{\pm,i}): \  ,
\end{multline}
where we set $\alpha_{(r_-,r_+)}\equiv r_+\sqrt{\beta}- r_-/\sqrt{\beta}-Q$. Then, we let $\mb{Z}_{(r-,r_+)}$ act on  external states and compute the normalized correlator 
\begin{multline}\label{corrQQ}
\frac{\bra{\alpha_\infty}\prod_A\mb{V}(y_A)\mb{Z}_{(r_-,r_+)}\prod_A\mb{V}(y_A/p)\ket{\alpha_0}}{\bra{0}\prod_A\mb{V}(y_A)\prod_A\mb{V}(y_A/p)\ket{0}}=\\
=(-p^{-1/2})^{r_+r_-}\,\oint\prod_\pm\prod_{i=1}^{r_\pm}\frac{\d z_{\pm,i}}{2\pi\i z_{\pm ,i}}\, z_{\pm ,i}^{\pm\sqrt{\beta^{\pm 1}}(\alpha_{(r_-,r_+)} +\alpha_0)}\times\\
\times\frac{\prod_\pm c_{\beta^{\pm 1}}(z_\pm;\mathfrak{q}_\pm)\,\Delta_{\mathfrak{q}_\mp^{-1}}(z_\pm;\mathfrak{q}_\pm)}{\prod_{i=1}^{r_+}\prod_{j=1}^{r_-}(1-p^{-1/2}z_{-,j}/z_{+,i})(1-p^{-1/2}z_{+,i}/z_{-,j})}\, \prod_\pm\prod_{i,A} \frac{(\mathfrak{q}_\pm^{1/2} y_A/p\,z_{\pm,i};\mathfrak{q}_\pm)_\infty}{(\mathfrak{q}_\pm^{1/2}  z_{\pm,i}/y_A ;\mathfrak{q}_\pm)_\infty}\ ,
\end{multline}
where $\alpha_\infty =\alpha_0+\alpha_{(2r_-,2r_+)}+Q$. If we set $\alpha_0\equiv \gamma_0-\alpha_{(r_-,r_+)}$, after suitable identifications, including 
\begin{align}
r_A&= r_{+A} \ ,  \quad c_A= r_{-A} \ , \quad r=r_+ =\sum_A r_{+A}\ , \quad c=r_-=\sum_A r_{-A}\ ,\nn\\
z^{\textrm{ R,L}}&=z^{-1}_{\pm}  \ ,\quad y_A^{-1}=p^{-1}\eta_\pm\mathfrak{q}_\pm^{1/2}x_At^{-r_A}q^{c_A} \ ,\quad \eta^{\textrm{R,L}}=\eta_\pm \ ,\quad \zeta^{\textrm{R,L}} =-\sqrt{\beta^{\pm 1}}\gamma_0 \ ,
\end{align}
we can match, up to normalization factors (see also footnote \ref{cbeta}), the correlator (\ref{corrQQ}) with the partition function (\ref{matrix-model}). Here, the decomposition $r_\pm =\sum_A r_{\pm,A}$ encodes a choice of integration contour, namely how the screening currents are distributed among the vertex operators. We refer to \cite{Aganagic:2013tta} and appendix \ref{KPapp} for more details. Finally, since we are considering an arbitrary number of screening charges, one can try to package all the correlators into a formal generating series
\begin{multline}
Z=\sum_{\vec r_\pm}K_{(\vec r_+,\vec r_-)}(\vec y,\gamma_0)\times\\
\times\frac{\bra{\gamma_0+Q}\e^{-\alpha_{(r_-,r_+)}\mb{Q}/2}\prod_A\mb{V}(y_A)\mb{Z}_{(r_-,r_+)}\prod_A\mb{V}(y_A/p)\, \e^{-\alpha_{(r_-,r_+)}\mb{Q}/2}\ket{\gamma_0}}{\bra{0}\prod_A\mb{V}(y_A)\prod_A\mb{V}(y_A/p)\ket{0}}\ ,
\end{multline}
where $K_{(\vec r_+,\vec r_-)}$ are suitable coefficients, which can be fixed so that $Z=Z_\text{inst}$. This example of BPS/CFT correspondence interpolates between the $q$-Virasoro/Vortex duality reviewed in this section and the $q$-Virasoro/Instanton duality reviewed in appendix \ref{KPapp}.

\section{Discussion}

In this note, we have proposed a set of new expansions of the instanton partition function of  5d $\mathcal{N}=1$ $\textrm{U}(N)$ pure Yang-Mills  theory, labeled by a choice of integers $\mathfrak{n} \in \mathbb{Z}^N$. The summands of these expansions admit an elegant interpretation in terms of 3d $\mathcal{N}=2$ partition functions of unitary gauge theories on $\union$ seen as a self-intersecting subspace of $\mathbb{C}_{q,t^{-1}}^2 \times \mathbb{S}^1$. Following and generalizing the work in \cite{Aganagic:2013tta,Kimura:2015rgi}, we have also given the $q$-Virasoro free boson realization of these new expansions, in terms of the two types of screening charges. As mentioned in the introduction, similar results can be obtained for the 4d reduction and the 6d lift on the torus, in which case the lower dimensional theories live on $(\mathbb{C}_{q})\cup (\mathbb{C}_{t^{-1}})$ and $(\mathbb{C}_{q}\times\mathbb{T}^2)\cup (\mathbb{C}_{t^{-1}}\times\mathbb{T}^2)$ respectively. From the algebraic perspective, the $q$-Virasoro algebra is replaced by its additive \cite{Hou:1996fx} or elliptic counterparts \cite{Nieri:2015dts,Kimura:2016dys}.  

It is straightforward to include fundamental hyper multiplets into the instanton partition function and derive the corresponding new expansions, as the building blocks are precisely $N_{Y_A \emptyset}$ and $N_{\emptyset Y_A}$ which also admit fairly simple factorization similar to (\ref{main-claim}). The resulting 3d partition functions will then have additional fundamental/anti-fundamental chiral multiplets. One can also generalize the analysis to other 5d unitary quiver gauge theories/W$_{q,t}$ algebras and to other systems coupled to codimension 2 and 4 BPS defects. For example, starting from a 5d linear quiver gauge theory one has a sum over Young diagrams for each gauge node, and therefore the Nekrasov partition function is of the form $\sum_{\vec Y_1} \sum_{\vec Y_2} \ldots $ with some intricate summand enjoying factorization properties similar to (\ref{main-claim}). One can then iteratively expand each sum $\sum_{\vec Y_k}$ one after another, where each step removes one 5d gauge node, but add one 3d gauge node to the resulting 3d left/right theories. As intermediate stages one gets the new expansions in terms of indices of 5d/3d/1d coupled systems. Ultimately one ends up with an expansion in terms of indices of 3d/1d coupled systems, where the left/right 3d theories are linear unitary quivers coupled through a collection of 1d chiral and Fermi multiplets. The detail for these cases is however beyond the scope of this paper. There are also conjectures \cite{Mitev:2014isa} of instanton partition functions for non-Lagrangian $T_N$ theories obtained by the method of topological vertex, and it would be very interesting to explore if they also admit similar 3d expansions and free boson realizations.

As discussed in \cite{Qiu:2014oqa}, multiple copies of 5d Nekrasov partition functions can be glued into 5d partition functions on compact toric Sasaki-Einstein manifolds. Therefore, we expect the expansions discussed in this note will have natural extensions to compact spaces. The $\mathbb{S}^5$ case is currently under investigation \cite{FYM}, and the relevant algebraic setting provided by the \textit{$q$-Virasoro modular triple} has recently been constructed in \cite{Nieri:2017vrb} (see also \cite{Nieri:2013vba} for earlier work in the context of 5d AGT).

So far, the new expansions that we propose lack a physical explanation or a first principle derivation. At the moment, we can only speculate that they correspond to some novel localization scheme. One might want to associate our results to switching off one non-commutative deformation \cite{Nekrasov:2013xda} in regularizing the instanton counting computation, as a consequence leading to 3d gauge theories on one $\mathbb{C} \times \mathbb{S}^1$. However, the fact that our expansions involve 3d gauge theories on the union $\union$ suggests that the physical origin is not of this nature. Another candidate derivation is the so-called  ``Higgs branch localization'' scheme \cite{Benini:2012ui,Benini:2013yva,Fujitsuka:2013fga,Doroud:2012xw,Peelaers:2014ima,Pan:2014bwa,Chen:2015fta,Pan:2015hza}, which localizes the path integral using certain well-chosen $\mathcal{Q}$-exact deformation term. Indeed, our result (\ref{final-expansion}) looks rather similar to those of the Higgs branch localization computation, where the matrix models are rewritten as sum of residues which can be organized into (products of) partition functions of infinitely many different theories, such as vortex/SW-partition functions. Moreover, the associated BPS configurations in 4d $\mathcal{N} = 2$ SQCD are shown to concentrate along intersecting $\mathbb{S}^2_\text{L} \cup \mathbb{S}^2_\text{R}$ in $\mathbb{S}^4_b$ \cite{Pan:2015hza}, which also leads to factorization of instanton partition functions similar to (\ref{main-claim}) in certain limit of the parameters $x_A$ \cite{Pan:2016fbl}. However, the Higgs branch localization requires the presence of fundamental matters, while the expansions we propose are valid without this limitation. Nevertheless, it is not unconceivable that some cleverly designed $\mathcal{Q}$-exact deformation term could lead to what we propose. Mathematically, these partial and alternative localization procedures might be related to equivariant localization on \textit{sub-strata} \cite{Atiyahbook}. If this is correct, then one should be able to identify the 3d gauge theory partition functions with some interesting equivariant cohomological quantity. Related to this possibility, it would be interesting to explore the relation (if any) between the subject addressed in this note and the categorification of complex Chern-Simons from 5d gauge theories as recently put forward in \cite{Gukov:2016gkn,Gukov:2017kmk} .

\acknowledgments
We thank N. Nekrasov, V. Pestun, J. Qiu, S. Shakirov and C. Vafa for valuable comments and discussions. We also thank the Simons Center for Geometry and Physics (Stony Brook University) for hospitality during the Summer Workshop 2017, at which some of the research for this paper was performed. The research of the authors is supported in part by Vetenskapsr\r{a}det under grant W2014-5517, by the STINT grant and by the grant ``Geometry and Physics" from the Knut and Alice Wallenberg foundation.

\appendix
\section{Special functions}\label{app:special-functions}

\subsubsection*{\texorpdfstring{$q$}{q}-Pochhammer symbols}

In this note we use the $q$-Pochhammer symbols  $(x;q)_\infty$ and $(x;q)_k$ extensively. They are defined by (when $|q| < 1$)
\begin{align}
  (x; q)_\infty = \prod_{i=0}^{+\infty}(1-x q^i)_\infty, \qquad (x;q)_k = \frac{(x;q)_\infty}{(xq^k;q)_\infty}\ , \quad \text{for }k \in \mathbb{Z} \ .
\end{align}
More explicitly,
\begin{align}
  (x;q)_k = \prod_{i=0}^{k-1} (1-x q^i), \quad \text{when } k\ge 0, \qquad (x;q)_k = \frac{1}{\prod_{i = 1}^{-k} (1 - x q^{-i})}, \quad \text{when }  k < 0 \ .
\end{align}
The $q$-Pochhammer symbol $(x;q)_\infty$ also admits a useful representation
\begin{align}
  (x; q)_\infty = \exp\Bigg[ - \sum_{m > 0} \frac{x^m}{m(1-q^m)} \Bigg] \ .
\end{align}

The symbol $(x;q)_k$ satisfies useful identities, among others
\begin{align}
  (x;q)_{n+k}=(x;q)_n (x q^n;q)_k \ . \label{qPoch-properties}
\end{align}

\subsection*{\texorpdfstring{$\Delta_t(z;q)$}{} and \texorpdfstring{$V_t(z,u;q)$}{}  }

In reorganizing the summands of the instanton partition functions, we define certain useful combinations of $q$-Pochhammer symbols which have gauge-theoretic as well as algebraic meaning.

The function $\Delta_t(z;q)$ is defined for a collection of $z = \{z_a\}$ variables as the product
\begin{align}
  \Delta_t(z;q) \equiv \prod_{a \ne b} \frac{(z_a /z_b;q)_\infty}{(t z_a /z_b ;q)_\infty} \ .
\end{align}
This is the Macdonald measure. In concrete situations, the collection $z$ can be as simple as $z = \{z_a| a = 1, \ldots, r\}$, or more involved ones like $z = \{z_{Ai} | A = 1, \ldots, N, \text{and } i = 1, \ldots, r_A\}$ and so forth. In the latter situation, we define
\begin{align}\label{Deltat}
  \Delta_t(z;q) = \prod_{A,B = 1}^N \prod_{i=1}^{r_A} \prod_{j=1}^{r_B} \Bigg|_{(A,i) \ne (B,j)} \frac{(z_{Ai} /z_{Bj};q)_\infty}{(t z_{Ai} /z_{Bj};q)_\infty} \equiv \prod_{(A,i) \ne (B,j)}\frac{(z_{Ai} /z_{Bj};q)_\infty}{(t z_{Ai} /z_{Bj};q)_\infty} \ .
\end{align}
The function $V_t(z,u;q)$ is defined in a similar spirit, namely
\begin{align}\label{Vt}
  V_t(z,u;q) \equiv \prod_a \frac{(z_a / u ;q)_\infty}{(t u/z_a;q)_\infty} \ .
\end{align}

\section{Derivations}\label{app:computations}

In this appendix, we collect the detailed derivation of the claim (\ref{main-claim}) in main text. The summands of the pure 5d $\textrm{U}(N)$ Yang-Mills instanton partition function can be written in terms of the Nekrasov function $N_{YW}$, which has the convenient product representation\footnote{We refer to \cite{Awata:2008ed} for more details and useful properties.}
\begin{align}
  N_{Y_A Y_B}(x;q,t) = & \prod_{i, j=1}^{+\infty} \frac{(x t^{j - i};q)_{Y_{Ai} - Y_{Bj}}}{(x t^{j - i + 1};q)_{Y_{Ai} - Y_{Bj}}} \ . \label{NYW}
\end{align}
To proceed, we follow the prescription in section \ref{subsection:new-expansion} and fix a difference vector $\vec{\mathfrak{n}}$. We extract for each Young diagram $Y_A$ its maximal rectangle, and denote the number of rows and columns of the rectangle to be $r_A, c_A$ respectively. Note that we have the inequities
\begin{align}
  Y_{Ai} \ge c_A, \quad i = 1, \ldots, r_A, \qquad \text{and} \qquad Y_{Ai} \le c_A, \quad i = r_A + 1, \ldots \ .
\end{align}
We can decompose the Young diagrams $Y_A$ into $Y^\text{L}_A$ and $Y^\text{R}_A$ as detailed in section \ref{subsection:new-expansion}
\begin{align}
  Y_{Ai} \equiv Y^\text{R}_{Ai}+c_A~,\quad i = 1, ..., r_A ~,\quad Y_{Ai} \equiv Y^\text{L}_{A,r_A + i}~,\quad i = 1, ..., +\infty~.
\end{align}

Now we are ready to factorize $N_{Y_A Y_B}$. By straightforward computation using (\ref{NYW}) and (\ref{qPoch-properties}) and the definition of $Y^\text{L}, Y^\text{R}$, we have
\begin{multline}
N_{Y_A Y_B}(x;q,t) =  \ \frac{N_{Y^\text{L}_A Y^\text{L}_B}(x t^{r_B-r_A};t^{-1},q^{-1})}{N_{Y^\text{L}_A \emptyset}(x t^{r_B-r_A}q^{-c_B};t^{-1},q^{-1})N_{\emptyset Y^\text{L}_B }(x t^{r_B-r_A}q^{c_A};t^{-1},q^{-1})}\times\\
  \times\frac{N_{Y^\text{R}_A Y^\text{R}_B}(x q^{c_A-c_B};q,t)}{N_{Y^\text{R}_A \emptyset}(x q^{c_A-c_B}t^{r_B};q,t)N_{\emptyset Y^\text{R}_B }(x q^{c_A-c_B}t^{-r_A};q,t)}\times\\
 \times \prod_{i=1}^{r_A}\prod_{j=1}^{r_B}\frac{(x t^{j-i};q)_{c_A-c_B}}{(t\; x t^{j-i};q)_{c_A-c_B}}\prod_{i=1}^{r_A}\prod_{j=1}^{\infty}\frac{(x t^{r_B} t^{j-i};q)_{c_A}}{(t\; x t^{r_B} t^{j-i};q)_{c_A}}\prod_{j=1}^{r_B}\prod_{i=1}^{\infty}\frac{(x t^{-r_A} t^{j-i};q)_{-c_B}}{(t\; x t^{-r_A} t^{j-i};q)_{-c_B}}\times\\
 \times N_{Y^\text{R}_A Y^{\text{L}\vee}_B}(x t^{r_B}q^{c_A};q,t)N_{Y^{\text{L}\vee}_A Y^\text{R}_B}(x t^{-r_A}q^{-c_B};q,t)~. \end{multline}
In the above, we renamed $Y^\text{L}\to Y^{\text{L}\vee}$ so that the new $Y^\text{L}_A$ has at most $c_A$ rows (instead of $c_A$ columns). We also applied the symmetry $N_{Y^\vee W^\vee}(x;q,t)=N_{Y W}(x;t^{-1},q^{-1})$ to the first line.

We notice that for Young diagrams $Y_A$ with at most $r_A$ rows, we can simplify the ratio
\begin{align}
    \frac{N_{Y_A Y_B}(x;q,t)}{N_{Y_A\emptyset}(x t^{r_B};q,t)N_{\emptyset Y_B}(x t^{-r_A};q,t)}=\prod_{i=1}^{r_A}\prod_{j=1}^{r_B}\frac{(t\; x q^{y_{Ai}-y_{Bj}}t^{j-i};q)_\infty}{(x q^{y_{Ai}-y_{Bj}} t^{j-i};q)_\infty}\frac{(x t^{j-i};q)_\infty}{(t\; x t^{j-i};q)_\infty}\ .
\end{align}
This simplification can be applied to the factors in the first and second row involving $Y^\text{L}$, $Y^\text{R}$ having at most $c_A$ and $r_A$ respectively. Therefore, $N_{Y_A Y_B}$ is reorganized into ratios of $q$-Pochhammer symbols, namely
\begin{multline}
  N_{Y_A Y_B}(x;q,t) =   \ \prod_{i=1}^{r_A}\prod_{j=1}^{r_B}\frac{(t\; x q^{c_A-c_B} q^{Y^\text{R}_{Ai}-Y^\text{R}_{Bj}}t^{j-i};q)_\infty}{(x q^{c_A-c_B} q^{Y^\text{R}_{Ai}-Y^\text{R}_{Bj}} t^{j-i};q)_\infty}\frac{(x q^{c_A-c_B} t^{j-i};q)_\infty}{(t\; x q^{c_A-c_B} t^{j-i};q)_\infty}\times\\
   \times \prod_{i=1}^{c_A}\prod_{j=1}^{c_B}\frac{(q^{-1}\; x t^{-(r_A-r_B)} t^{-(Y^\text{L}_{Ai}-Y^\text{L}_{Bj})}q^{-(j-i)};t^{-1})_\infty}{(x t^{-(r_A-r_B)} t^{-(Y^\text{L}_{Ai}-Y^\text{L}_{Bj})} q^{-(j-i)};t^{-1})_\infty}\frac{(x t^{-(r_A-r_B)} q^{-(j-i)};t^{-1})_\infty}{(q^{-1}\; x t^{-(r_A-r_B)} q^{-(j-i)};t^{-1})_\infty} \times\\
 \times \prod_{i=1}^{r_A}\prod_{j=1}^{r_B}\frac{(x t^{j-i};q)_{c_A-c_B}}{(x t^{1 + j-i};q)_{c_A-c_B}}\prod_{i=1}^{r_A}\prod_{j=1}^{\infty}\frac{(x t^{r_B} t^{j-i};q)_{c_A}}{(x t^{1+ r_B} t^{j-i};q)_{c_A}}\prod_{j=1}^{r_B}\prod_{i=1}^{\infty}\frac{(x t^{-r_A} t^{j-i};q)_{-c_B}}{( x t^{1-r_A} t^{j-i};q)_{-c_B}} \times\\
 \times N_{Y^\text{R}_A Y^{\text{L}\vee}_B}(x t^{r_B}q^{c_A};q,t)N_{Y^{\text{L}\vee}_A Y^\text{R}_B}(x t^{-r_A}q^{-c_B};q,t)~.
\end{multline}
We can now use another representation of $N_{YW}$, that is
\begin{align}
  N_{Y W}(x;q,t)=\prod_{i,j=1}^\infty\frac{1-x q^{Y _i-j}t^{W^\vee_j-i+1}}{1-x q^{-j}t^{-i+1}}~,
\end{align}
to reorganize the factors in the last line by unpacking the product over $i, j = 1, \ldots, +\infty$ to $i, j = 1, \ldots, r_A (c_A)$ and $r_A + 1 (c_A + 1),  \ldots , +\infty$, namely
\begin{multline}
  N_{Y^{\text{L}\vee}_A Y^\text{R}_B}(x t^{-r_A}q^{-c_B};q,t) =  \ N_{Y^\text{L}_A Y^{R\vee}_B}(x t^{-r_A}q^{-c_B};t^{-1},q^{-1}) =\\
  % = & \ \prod_{i=1}^{c_A}\prod_{j=1}^{r_B}\frac{1-q^{-1}x t^{-r_A+j}q^{-c_B+i}t^{-Y^\text{L}_{Ai}}q^{-Y^\text{R}_{Bj}}}{1-q^{-1}x t^{-r_A+j}q^{-c_B+i}}\\
  % & \times \prod_{i=1}^{c_A}\frac{(t q^{-1} x t^{r_B-r_A}q^{-c_B+i} t^{-Y^\text{L}_{Ai}};t)}{(t q^{-1} x t^{r_B-r_A}q^{-c_B+i};t)}\prod_{j=1}^{r_B}\frac{(x t^{-r_A+j}q^{c_A-c_B} q^{-Y^\text{R}_{Bj}};q)}{(x t^{-r_A+j}q^{c_A-c_B};q)}\\
  =  \prod_{i=1}^{c_A}\prod_{j=1}^{r_B}\frac{1-q^{-1}x t^{-r_A+j}q^{-c_B+i}t^{-Y^\text{L}_{Ai}}q^{-Y^\text{R}_{Bj}}}{1-q^{-1}x t^{-r_A+j}q^{-c_B+i}}  \times \\
  \times \prod_{i=1}^{c_A}\frac{(q^{-1} x t^{r_B-r_A}q^{-c_B+i};t^{-1})}{(q^{-1} x t^{r_B-r_A}q^{-c_B+i} t^{-Y^\text{L}_{Ai}};t^{-1})}\prod_{j=1}^{r_B}\frac{(x t^{-r_A+j}q^{c_A-c_B} q^{-Y^\text{R}_{Bj}};q)}{(x t^{-r_A+j}q^{c_A-c_B};q)}~,
\end{multline}
and
\begin{align}
  N_{Y^\text{R}_A Y^{\text{L}\vee}_B}(x t^{r_B}q^{c_A};q,t) = & \ \prod_{i=1}^{r_A}\prod_{j=1}^{c_B}\frac{1-t x t^{r_B-i}q^{c_A-j}q^{Y^\text{R}_{Ai}}t^{Y^\text{L}_{Bj}}}{1-t x t^{r_B-i}q^{c_A-j}} \nonumber\\
  % = \prod_{i=1}^{r_A}\prod_{j=1}^{c_B}\frac{1-t x t^{r_B-i}q^{c_A-j}q^{Y^\text{R}_{Ai}}t^{Y^\text{L}_{Bj}}}{1-t x t^{r_B-i}q^{c_A-j}}\\
  % \times \prod_{i=1}^{r_A}\frac{(t q^{-1} x t^{r_B-i} q^{c_A-c_B}q^{Y^\text{R}_{Ai}};q^{-1})}{(t q^{-1} x t^{r_B-i} q^{c_A-c_B};q^{-1})}\prod_{j=1}^{c_B}\frac{(x t^{r_B-r_A} q^{c_A-j} t^{Y^\text{L}_{Bj}};t^{-1})}{(x t^{r_B-r_A} q^{c_A-j};t^{-1})}=\\
  & \times \prod_{i=1}^{r_A}\frac{(t x t^{r_B-i} q^{c_A-c_B};q)_\infty}{(t  x t^{r_B-i} q^{c_A-c_B}q^{Y^\text{R}_{Ai}};q)_\infty}\prod_{j=1}^{c_B}\frac{(x t^{r_B-r_A} q^{c_A-j} t^{Y^\text{L}_{Bj}};t^{-1})_\infty}{(x t^{r_B-r_A} q^{c_A-j};t^{-1})_\infty}~.
\end{align}
Now we set $x=x_{AB}=x_A /x_B$, and define $w_{Y_{Ai}^\text{R}} \equiv x_A q^{c_A}q^{Y^\text{R}_{Ai}}t^{1-i}$, $z_{Y_{Ai}^\text{L}} \equiv x_A t^{-r_A}t^{-Y^\text{L}_{Ai}}q^{i-1}$. Similarly, we define $w_{\emptyset^\text{R}_{Ai}} \equiv x_A q^{c_A}t^{1-i}$, $z_{\emptyset^\text{L}_{Ai}} \equiv x_A t^{- r_A}q^{i-1}$. With these new variables, we observe that various combinations of $x, q, t$ in $N_{Y_A Y_B}$ organize into ratios
\begin{align}
  \frac{w_{Y^\text{R}_{Ai}}}{w_{Y^\text{R}_{Bj}}} = x_{AB}q^{c_A - c_B}q^{Y^\text{R}_{Ai} - Y^\text{R}_{Bj} }t^{j-i}, \qquad \frac{z_{Y^\text{L}_{Ai}}}{z_{Y^\text{L}_{Bj}}} = x_{AB}t^{- (r_A - r_B)}t^{- (Y^\text{L}_{Ai} - Y^\text{L}_{Bj}) }q^{- (j-i)} \ ,
\end{align}
and their $Y \to \emptyset$ counterparts. Now we can take the product over $A, B = 1, \ldots, N$, and rename some of the $(A,i), (B,j)$ indices. We end up with
\begin{multline}
  \prod_{A,B}N_{Y_A Y_B}(x_{AB};q,t) =\\
  =  \frac{\Delta_t(\vec w_{\emptyset^\text{R}};q)}{\Delta_t(\vec w_{\vec Y^\text{R}};q)} \prod_{A,B=1}^N\prod_{j=1}^{r_A}\frac{(t (w_{Y^\text{R}_{Aj}})^{-1} x_B t^{-r_B} q^{c_B};q)_\infty}{(t (w_{\emptyset^\text{R}_{Aj}})^{-1} x_B t^{-r_B} q^{c_B};q)_\infty}\frac{(w_{\emptyset^\text{R}_{Aj}}x_B^{-1} t^{r_B}q^{-c_B};q)_\infty}{(w_{Y^\text{R}_{Aj}}x_B^{-1} t^{r_B}q^{-c_B};q)_\infty} \times\\
 \times \frac{\Delta_{q^{-1}}(\vec z_{\emptyset^\text{L}};t^{-1})}{\Delta_{q^{-1}}(\vec z_{\vec Y^\text{L}};t^{-1})}\prod_{A,B}\prod_{j=1}^{c_A}\frac{(q^{-1} (z_{Y^\text{R}_{Aj}})^{-1} x_b t^{-r_B} q^{c_B};t^{-1})_\infty}{(q^{-1}(z_{\emptyset^\text{R}_{Aj}})^{-1}x_B t^{-r_B} q^{c_B};t^{-1})_\infty}\frac{(z_{\emptyset^\text{L}_{Aj}}x_B^{-1} t^{r_B} q^{-c_B};t^{-1})_\infty}{(z_{Y^\text{L}_{Aj}}x_B^{-1} t^{r_B} q^{-c_B};t^{-1})_\infty} \times\\
 \times\prod_{A,B}\prod_{i=1}^{r_A}\prod_{j=1}^{c_B}\frac{1-t z_{Y^\text{L}_{Bj}} (w_{Y^\text{R}_{Ai}})^{-1}}{1-t z_{\emptyset^\text{L}_{Bj}} (w_{\emptyset^\text{R}_{Ai}})^{-1}}\frac{1-q^{-1} w_{Y^\text{R}_{Ai}}  (z_{Y^\text{L}_{Bj}})^{-1}}{1-q^{-1} w_{\emptyset^\text{R}_{Ai}} (z_{\emptyset^\text{L}_{Bj}})^{-1}} \times\\
 \times \prod_{A,B}\prod_{i=1}^{r_A}\frac{(x_{AB} t^{1-i};q)_{c_A-c_B}}{(x_{AB} t^{1+r_B-i};q)_{c_A-c_B}}\frac{(x_{AB} t^{r_B+1-i};q)_{c_A}}{(x_{AB}^{-1} t^{-r_B+i};q)_{-c_A}}~,
\end{multline}
where the last line come from
\begin{multline}
\prod_{A,B}\prod_{i=1}^{r_A}\prod_{j=1}^{r_B}\frac{(x_{AB} t^{j-i};q)_{c_A-c_B}}{(t\; x_{AB} t^{j-i};q)_{c_A-c_B}}\prod_{i=1}^{r_A}\prod_{j=1}^{\infty}\frac{(x_{AB} t^{r_B} t^{j-i};q)_{c_A}}{(t\; x_{AB} t^{r_B} t^{j-i};q)_{c_A}}\prod_{j=1}^{r_B}\prod_{i=1}^{\infty}\frac{(x_{AB} t^{-r_A} t^{j-i};q)_{-c_B}}{(t\; x_{AB} t^{-r_A} t^{j-i};q)_{-c_B}} =\\
  =  \prod_{A,B}\prod_{i=1}^{r_A}\frac{(x_{AB} t^{1-i};q)_{c_A-c_B}}{(x_{AB} t^{1+r_B-i};q)_{c_A-c_B}}\frac{(x_{AB} t^{r_B+1-i};q)_{c_A}}{(x_{AB}^{-1} t^{-r_B+i};q)_{-c_A}}~.
\end{multline}

Finally, we rescale all $w\to  w/\eta^{\textrm{R}}$, $z\to z/\eta^{\textrm{L}}$ with $\eta^{\textrm{L}}/\eta^{\textrm{R}}=(q t)^{1/2}$, so that we have
\begin{align}
  w_{Y_{Ai}^\textrm{R}}=\eta^\textrm{R} x_A q^{c_A}q^{Y^\text{R}_{Ai}}t^{1-i}~,\quad z_{Y_{Ai}^\text{L}}=\eta^\textrm{L}x_A t^{-r_A}t^{-Y^\text{L}_{Ai}}q^{i-1}~.
\end{align}
%and introduce
%\begin{align}
%  V(\vec w_{\vec Y^\text{L}},u;q,t)&=\prod_{A}\prod_{j=1}^{r_A}\frac{(w_{Y_{Aj}^\text{R}}\; u^{-1};q)}{(t\; (w_{Y_{Aj}^\text{R}})^{-1}\; u;q)}, \qquad V(\vec z_{\vec Y^\text{R}},v;t^{-1},q^{-1})=\prod_{A}\prod_{j=1}^{c_A}\frac{(z_{Y_{Aj}^\text{L}}\; v^{-1};t^{-1})}{(q^{-1}\; (z_{Y_{Aj}^\text{L}})^{-1}\; v; t^{-1})}~, \nonumber\\
%  V_{\rm int}(\vec w_{\vec y^{L}},\vec z_{\vec Y^\text{R}};p)&=\prod_{A,B}\prod_{i=1}^{r_A}\prod_{j=1}^{c_B}\frac{1}{(1-p^{-1/2} z_{Y^\text{L}_{Bj}} (w_{Y^\text{R}_{Ai}})^{-1})(1-p^{-1/2} w_{Y^\text{R}_{Ai}}(z_{Y^\text{L}_{Bj}})^{-1})}~.
%\end{align}
We then arrive at the final expression for the product $\prod_{A,B} N_{Y_A Y_B}$, that is
\begin{multline}
 \prod_{A,B}  N_{Y_A Y_B}(x_{AB};q,t) =  \frac{\Delta_t(\vec w_{\emptyset^R};q)}{\Delta_t(\vec w_{\vec Y^\text{R}};q)} \frac{\Delta_{q^{-1}}(\vec z_{\emptyset^L};t^{-1})}{\Delta_{q^{-1}}(\vec z_{\vec Y^\text{L}};t^{-1})}   \times\\
  \times      \prod_{B = 1}^N\frac{V(\vec w_{\vec \emptyset^\text{R}},\eta^{-1} x_B t^{-r_B} q^{c_B};q,t)}{V(\vec w_{\vec Y^\text{R}},\eta^{-1} x_B t^{-r_B} q^{c_B};q,t)}\prod_{B=1}^N\frac{V(\vec z_{\vec \emptyset^L},\xi^{-1}x_B t^{-r_B} q^{c_B};t^{-1},q^{-1})}{V(\vec z_{\vec Y^\text{L}},\xi^{-1}x_B t^{-r_B} q^{c_B};t^{-1},q^{-1})}\times\\
 \times \frac{V_{\rm int}(\vec w_{\vec \emptyset^\text{R}},\vec z_{\vec \emptyset^L};p)}{V_{\rm int}(\vec w_{\vec Y^\text{R}},\vec z_{\vec Y^\text{L}};p)} \prod_{A,B = 1}^N\prod_{i=1}^{r_A}\frac{(x_{AB} t^{1-i};q)_{c_A-c_B}}{(x_{AB} t^{1+r_B-i};q)_{c_A-c_B}}\frac{(x_{AB} t^{r_B+1-i};q)_{c_A}}{(x_{AB}^{-1} t^{-r_B+i};q)_{-c_A}}~,
\end{multline}
where the functions $\Delta$ and $V$ are defined in (\ref{Deltat}), (\ref{Vt}). We point out that the last factor dependents only the shape of the maximal rectangle, but not on the subdiagrams $Y^\text{L,R}_A$. This concludes the derivation of the claim (\ref{main-claim}).

\section{\texorpdfstring{Index on $\mathbb{C}_q\times \mathbb{S}^1$}{}\label{app:index}}

In this appendix we collect some relevant results from \cite{Yoshida:2014ssa}. The index of an $\mathcal{N} = 2$ $\textrm{U}(n)$ gauge theory with a collection of chiral multiplets with either Neumann or Dirichlet boundary conditions on the bulk $\mathbb{D}^2\times\mathbb{S}^1\simeq \mathbb{C}_q \times \mathbb{S}^1$, coupled with some 2d $\mathcal{N} = (0,2)$ multiplets on the boundary $\mathbb{T}^2_q\simeq \mathbb{S}^1 \times \mathbb{S}^1$, is given by\footnote{In the absence of any two dimensional boundary interaction, $I^{\mathbb{C}_q\times \mathbb{S}^1} = I^{\mathbb{D}^2_q \times \mathbb{S}^1}$.}
\begin{align}
  Z^{\mathbb{C}_q \times \mathbb{S}^1} =
    \int \d^n \sigma \,
      Z_\text{cl}(\sigma) Z^\text{3d}_\text{1-loop} (\sigma)Z^\text{2d}_\text{1-loop} (\sigma) \ .
\end{align}
The classical action receives contributions from mixed Chern-Simons terms. The 3d 1-loop determinant receives contributions from $\textrm{U}(n)$ vector multiplets and chiral multiplets transforming in different representations of $\textrm{U}(n)$ with Neumann or Dirichlet boundary conditions. Their flavor symmetries can be weakly gauged by background vector multiplets, therefore introducing real masses $\mu$. They also carries $\mathcal{R}$-charges $\Delta$. One can form complex masses by defining
\begin{align}
  \text{fundamental}: & \quad m \equiv \mu + \frac{\Delta \epsilon_1}{2} \ , \qquad \text{anti-fundamental}:  \quad \tilde m \equiv \tilde{\mu} - \frac{\tilde \Delta \epsilon_1}{2} \ , 
  \end{align}
  \begin{align}
  \text{adjoint}: & \quad \tilde m_{\textrm{ad}} \equiv \mu_\textrm{ad} + \frac{\Delta_{\textrm{ad}} \epsilon_1}{2} \ .
\end{align}
Here we defined $q \equiv \e^{2\pi \i \epsilon_1}$. The contributions to the 3d and 2d 1-loop determinants include the following:\footnote{We choose to ignore the exponential factors arising from regularization. We have rescaled and renamed the parameters by $\i \beta r \rho(a) \to 2\pi \i \rho(\sigma), \quad F_l M_l \to 2\pi \i F_A \mu_A, \quad e^{- 2 \beta_2} \to e^{2\pi \i \epsilon_1}$. We also adopt the quiver convention for the equivariant parameters, so that $N$ fundamental chiral multiplets transforms in the anti-fundamental of the $\textrm{U}(N)$ flavor group, with $F_A = - 1$. The resulting equivariant parameters will behave like $\rho(\sigma) + F_A \mu_A \to \sigma_a - \mu_A$.}
\begin{itemize}[leftmargin=*]
  \item vector multiplet contributes
  \begin{align}
    Z^{\mathbb{C}_q \times \mathbb{S}^1}_\text{vector}(\sigma) = \prod_{\substack{a, b = 1 \\ a\ne b}}^n
      (\e^{2\pi \i (\sigma_a - \sigma_b)};q)_\infty \ ;
  \end{align}
  \item $N$ chiral multiplets with Neumann (N) or Dirichlet (D) boundary conditions transforming in the representation $\rho$ of the $\textrm{U}(n)$ gauge group contribute
  \begin{align}
    Z^{\mathbb{C}_q \times \mathbb{S}^1}_\text{N} & = \prod_{A=1}^N\prod_{w\in \rho} \frac{1}{(\e^{ - 2\pi \i( w(\sigma) + F_A m_A) } ; q)_\infty} \ , \quad Z^{\mathbb{C}_q \times \mathbb{S}^1}_\text{D}  = \prod_{A=1}^N\prod_{w\in \rho} (q\, \e^{  2\pi \i( w(\sigma) + F_A m_A) } ; q)_\infty \ ,
  \end{align}
where $w$ denotes the weights in the representation $\rho$. %  \item $N_1$ fundamental and $N_2$ anti-fundamental chiral multiplets, with $\mathcal{R}$-charge $\Delta$ and Dirichlet boundary condition, contribute\footnote{In \cite{Yoshida:2014ssa}, there are typos in both equation (D.29) and (4.13). The correct result should follow from the line above (D.29), and reads $e^{- \mathcal{E}(\i \beta r \rho(a) + (\Delta - 2)\beta_2 + F_l M_l)}(e^{\i \beta r \rho(a) + F_l M_l}q^{2 - \Delta};q^2)$, with $q = e^{-\beta_2}$. In our current convention, it reads $e^{- \mathcal{E}(2\pi \i ( \pm \sigma_a \mp {m}_A + \epsilon_1)  )}(e^{ \pm 2\pi \i (\sigma_a - m_A)} q; q)$ for fundamental/anti-fundamental chiral multiplets, where we have absorbed the $\textrm{U}(1)_\mathcal{R}$ charge $\Delta$ in the complex masses.}
%  \begin{align}
%    Z^{\mathbb{C}_q \times \mathbb{S}^1}_\text{fund-D}Z^{\mathbb{C}_q \times \mathbb{S}^1}_\text{afund-D} = \prod_{a = 1}^n \prod_{A = 1}^N \prod_{w \in \mathcal{R}} (e^{2\pi \i(w(\sigma ) + F_A m_A + {\epsilon _1})};q)   \ .
%  \end{align}
%  \item One adjoint chiral multiplet of $\mathcal{R}$-charge $\Delta_X$ with either boundary condition contribute
%  \begin{align}
%    &  Z^{\mathbb{C}_q \times \mathbb{S}^1}_\text{adj-N} = \prod_{\substack{a,b=1 \\ a \ne b}}^n  \frac{1}{(e^{ - 2\pi \i( \sigma_a - \sigma_b - m_X) } ; q)} \, \\
%    &  Z^{\mathbb{C}_q \times \mathbb{S}^1}_\text{adj-D} = \prod_{\substack{a,b=1 \\ a \ne b}}^n  (e^{ + 2\pi \i( \sigma_a - \sigma_b - m_X) } q ; q)\ .
%  \end{align}
\item boundary multiplets contribute \cite{Gadde:2013ftv,Benini:2013nda,Benini:2013xpa}
\begin{align}
Z^{\mathbb{T}^2_q}_{\textrm{chiral}}&=\frac{1}{\Theta(\e^{-2\pi\i (w(\sigma)+\nu)};q)} \ , \quad Z^{\mathbb{T}^2_q}_{\textrm{Fermi}}=\Theta(\e^{2\pi\i (w(\sigma)+\nu)};q) \ ,
\end{align}
where $\nu$ is some $\textrm{U}(1)$ mass parameter. Notice that the 1-loop determinants of 3d chiral multiplets with opposite boundary conditions can be related using the identity
\be
(q\, \e^{2\pi\i(w(\sigma)+m)};q)_\infty=\frac{\Theta(\e^{-2\pi\i(w(\sigma)+m)};q)_\infty}{(\e^{-2\pi\i(w(\sigma)+m)};q)_\infty} \ .
\ee
This can be related to anomaly cancellation conditions of Chern-Simons terms in the presence of a boundary, and each $\Theta$ function is associated to a Chern-Simons unit. 
\end{itemize}

Let us examine the special case of a $\textrm{U}(n)$ gauge theory, coupled to 1 adjoint, $N$ fundamental and $N$ fundamental chiral multiplets, each with Neumann, Neumann and Dirichlet boundary condition respectively. In this case, the 1-loop determinant reads
\begin{align}
  Z_{1\textrm{-loop}}^{\mathbb{C}_q \times \mathbb{S}^1}= & \ \prod_{\substack{a,b=1 \\ a \ne b}}^n 
    \frac{(\e^{2\pi \i (\sigma_a - \sigma_b)};q)_\infty}{(\e^{ - 2\pi \i( \sigma_a - \sigma_b - m_{\text{ad}}) } ; q)_\infty}  \prod_{A=1}^N
    \frac{(q \ \e^{ + 2\pi \i( \sigma_a - m^\text{D}_A) } ; q)_\infty}{(\e^{ - 2\pi \i( \sigma_a - m^\text{N}_A) } ; q)_\infty} \ .
\end{align}

\section{Free sector}\label{app:freesec}

The products of $q$-Pochhammer symbols in the prefactor $1/\mathcal{B}N_{\square}$, as written in (\ref{prefactor}), can also be recognized as the partition function of a collection of free chiral multiplets on $\mathbb{C}_q \times \mathbb{S}^1$ and $\mathbb{C}_{t^{-1}} \times \mathbb{S}^1$, together with a collection of 1d free chiral and Fermi multiplets on the intersection $\mathbb{S}^1$
\begin{align}
  \frac{Q^{\vec r \cdot \vec c}}{\mathcal{B}(\vec r, \vec c)N_{\square}(\vec r, \vec c)} = & \  W_{\vec r,\vec c} \, Z^{\mathbb{C}_{t^{-1}} \times \mathbb{S}^1}_\text{free}  Z^{\mathbb{C}_q \times \mathbb{S}^1}_\text{free}  Z^{\mathbb{S}^1}_\text{free chiral+Fermi} \ .
\end{align}
Here, $Z^{\mathbb{C}_q \times \mathbb{S}^1}$ receives contributions from two sets of Neumann and two sets of Dirichlet free chiral multiplets, with masses $m_{A, Bi}$ listed in the following table ($i = 1, \ldots, r_B$):
\begin{center}
  \begin{tabular}{c|c|c}
    & Neumann                            & Dirichlet \\ \hline
   $m^\text{R}_{A,Bi}$& $ - (X_A + r_A \epsilon_2) + X_B + (i-1) \epsilon_2$    & $-X_A + X_B + \epsilon_1 + i \epsilon_2$
  \end{tabular} \ .
\end{center}
Similarly for $Z^{\mathbb{C}_{t^{-1}} \times \mathbb{S}^1}$, with replacement $r_A \leftrightarrow c_A$, $\epsilon_1 \leftrightarrow \epsilon_2$. These free 3d chiral multiplets organize into bi-fundamental representations of some $\textrm{U}(N) \times \textrm{U}(r)$ flavor group(s). The 1d term $Z^{\mathbb{S}^1}_\text{free chiral+Fermi}$ receives contributions from two sets of free Fermi and two sets of free chiral multiplets, with masses listed in the following table:
\begin{center}
  \begin{tabular}{c|c}
    Fermi $m_{A,Bij}$ & chiral $m_{Ai, Bj}$\\
    $i = 1, \ldots, r_B$, $ j = 1, \ldots, c_B$ & $i = 1, \ldots, r_A$, $ j = 1, \ldots, c_B$\\\hline
    $M_A - M_{Bij}$ &  $M^{\epsilon_2}_{Ai} - (M_{Bj}^{\epsilon_1} + r_B \epsilon_2)$\\
    $M_A + c_A \epsilon_1 + r_A \epsilon_2 - M_{Bij}$ & $(M_{Ai}^{\epsilon_2} + c_A \epsilon_1) - M_{Bj}^{\epsilon_1}$
  \end{tabular} \ ,
\end{center}
where we have defined the equivariant mass parameters
\begin{align}
  M_A \equiv X_A, \quad M_{Ai}^\epsilon \equiv X_A + (i - \frac{1}{2})\epsilon, \quad M_{Bij} \equiv X_B + (j-\frac{1}{2})\epsilon_1 + (i - \frac{1}{2})\epsilon_2 \ .
\end{align}
As indicated by the names of the masses, the Fermi multiplets organize into bi-fundamental representations of some $\textrm{U}(N) \times \textrm{U}(\vec r \cdot \vec c)$ flavor symmetry group(s), while the chiral multiplets organize into bi-fundamental representation of some $\textrm{U}(r) \times \textrm{U}(c)$ flavor group(s).\footnote{We note that there are different equivariant mass parameters, which correspond to different flavor symmetry groups. For instance, the $\textrm{U}(N)$ parameters $M_A$ and $M_A + c_A \epsilon_1 + r_A \epsilon_2$ correspond to different $\textrm{U}(N)$ symmetries.}
% It is straightforward to work out the exponential factor $W_\text{free}(m)$
% \begin{align}
%   W_\text{free}(m) 
% %  & \ \Bigg[ \prod_{A,B=1}^N \prod_{i=1}^{r_B}\frac{E\big(2\pi \i (- m_{A,Bi}^\text{R,D} + \epsilon_1)\big)}{E\big(2\pi \i (- m_{A,Bi}^\text{R,N})\big)} \Bigg] \Bigg[\epsilon_1 \leftrightarrow \epsilon_2, r \leftrightarrow c\Bigg] \nonumber\\
%   = & \ \Bigg[\e^ {\sum_{A,B=1}^N\sum_{i=1}^{r_B}\frac{\pi \i}{2\epsilon _1}(m_{A,Bi}^\text{R,D} + m^\text{R,N}_{A,Bi} - 2\epsilon _1)(\epsilon _1 + m^\text{R,N}_{A,Bi} - m^\text{R,D}_{A,Bi})}\Bigg]  \Bigg[\epsilon_1 \leftrightarrow \epsilon_2, r \leftrightarrow c\Bigg] \ . \label{free-exponential-factors}
% \end{align}
Finally, the coefficient $W_{r,c} (m)$ reads
\begin{align}
  W_{\vec r,\vec c} \equiv & \ \frac{{\eta ^{-r{\zeta ^{\text{R}}}}}{\eta ^{-c{\zeta ^{\text{L}}}}}}{Q_g^{\vec r \cdot \vec c}} \Bigg[ \prod_{A = 1}^N \prod_{i=1}^{r_A} (x_A t^{1-i} )^{-\zeta^\text{R}}\Bigg]\Bigg[\prod_{A = 1}^N \prod_{j=1}^{c_A} ( x_A q^{j-1} )^{-\zeta^\text{L}}\Bigg] \nonumber\\
  & \ \times \Bigg[ \frac{1}{(t;q)_\infty} \Res_{z \to 1} \frac{1}{z(z^{-1};q)_\infty} \Bigg]^r \Bigg[\frac{1}{(q^{-1};t^{-1})_\infty} \Res_{z \to 1} \frac{1}{z(z^{-1};t^{-1})_\infty} \Bigg]^c \ .
\end{align}

\section{Infinitely-many screening charges}\label{KPapp}
Let us consider \textit{based} screening charges defined by Jackson integrals \cite{Kimura:2015rgi}, namely 
\be\label{Qjack}
\mb{Q}^{(\pm)}_{z}\equiv \sum_{k\in\mathbb{Z}}z \mathfrak{q}_\pm^k \ \mb{S}^{(\pm)}(z \mathfrak{q}_\pm^k)\ .
\ee
This (less familiar) definition allows one to consider the insertion of infinitely-many screening charges as there are no explicit integrals to compute, and  an additional label attached to the screening charge as the base point $z$ is quite a free parameter. Therefore, we can consider infinitely-many base points in the set
\be
\chi_\emptyset \equiv \{x_{Ai}\equiv x_A t^{1-i}\ | A = 1, ..., N \ , i = 1,..., \infty\} \ ,
\ee
and define the operator 
\be
\mb{Z}\equiv  \prod^{\succ}_{z\in \chi_\emptyset}\mb{Q}^{(+)}_z  \ ,
\ee
where $\prod^\succ$ denotes an ordered product\footnote{\label{ord}We define the order $\succ$ on $\chi_\emptyset$ by declaring $x_{Ai}\succ x_{Bj}$ if $A>B$, and for $A=B$ if $i\geq j$. The ordered product $\prod^\succ$ follows the reverse ordering}. Notice that we have again explicitly broken the $q\leftrightarrow t^{-1}$ symmetry by considering only one kind of screening charge and a specific set of base points. However, this symmetry will be at the end restored by the infinite product. In order to recast this state in a more familiar form, one  observe that the points $x_{A}t^{1-i}q^{k_{Ai}}$ give rise to zeros in the ``OPE" function of the screening charges, unless they fall into a Young diagram classification, namely $k_{Ai}\ge k_{A,i+1} \ge 0$. Therefore, we denote the set of contributing points as (now replacing $k_A$ with Young diagrams $Y_A$)
\be
\chi\equiv \{ x_{Y_{Ai}}\equiv x_A t^{1-i} q^{Y_{Ai}}\ | A = 1, ..., N \ , i = 1,..., \infty\} \ ,
\ee
where $\vec Y\equiv (Y_1,\ldots, Y_N)$ is a collection of Young diagrams, and write\footnote{Since we are dealing with infinite products, some care with regularization is needed. In this note, we do not address this issue in detail but we simply observe that some divergence can be reabsorbed into $\mu_0$, which has in fact to ``absorb" an infinite number of screening charges.}
\be
\mb{Z}= \sum_{\vec Y} \prod^{\succ}_{z\in \chi}z\ \mb{S}^{(+)}(z)\ .
\ee
Proceeding formally as in the finite case, we can write
\be
\mb{Z}=\widehat c_\beta( x_\emptyset;q)\sum_{\vec Y} \Big( \widehat{\Delta}_t(x_Y;q)\ \prod_{z\in\chi}^\succ z^{\sqrt{\beta}(\sqrt{\beta}|\chi|-Q)}:\prod_{z\in\chi}^{\succ}\mb{S}^{(+)}(z): \Big)\ ,
\ee
where $x_\emptyset\in\chi_\emptyset$, $x_Y\in\chi$, and the hat reminds us that we are considering infinitely-many variables (the affine limit). With an abuse of notation, we have denoted by $|\chi|$ the (infinite) number of screening charges. Now we  notice that
\be
\widehat{\Delta}_t( x_Y;q)=\prod_{\substack{(A,i)\neq (B,j)\\ A,B = 1,...,N\\ i,j = 1,...,\infty}}\frac{(x_{AB}q^{Y_{Ai}-Y_{Bj}}t^{j-i};q)_\infty}{(t x_{AB}q^{Y_{Ai}-Y_{Bj}}t^{j-i};q)_\infty}=\frac{\widehat{\Delta}_t(x_\emptyset;q,t)}{\prod_{A,B=1}^N N_{Y_A Y_B}(x_{AB};q,t)}~.
\ee
Therefore, we compute the properly (re)normalized correlator
\be
\frac{\bra{\mu_\infty}\mb{Z}\ket{\mu_0}}{\bra{\mu_\infty}\prod^{\succ}_{z\in\chi_\emptyset} z \ \mb{S}^{(+)}(z)\ket{\mu_0}}=\sum_{\vec Y}\ \frac{Q_g^{|\vec Y|}}{\prod_{A,B=1}^N N_{Y_A Y_B}(x_{AB};q,t)} \ ,
\ee
where the external states are eigenstates of $\mb{P}$ and $\mu_\infty$ is chosen to ensure charge conservation, with $Q_g\equiv q^{\sqrt{\beta}(\sqrt{\beta}|\chi|-Q+\mu_0)}$. As follows from the BPS/CFT correspondence, in the expression above we can easily recognize the Nekrasov instanton partition function of 5d $\textrm{U}(N)$ pure Yang-Mills theory. Finally, the inclusion of an equal number of fundamental and anti-fundamental matter is equivalent to the normalized correlator
%\footnote{Notice that the identification $
%\mb{a}_{-m}\simeq -\frac{q^{m/2}}{m} (Q_f^m-\bar Q_f^{-m})$ would lead to a pair of fundamentals, which can be related to a fundamental/anti-fundamental pair by a Chern-Simons unit.}
\begin{multline}\label{VZVinst}
\frac{\bra{\mu_\infty}\prod_f \mb{V}(Q_f)\mb{Z}\prod_f \mb{V}(\bar Q_f)\ket{\mu_0}}{\bra{\mu_\infty}\prod_f \mb{V}(Q_f)\prod^{\succ}_{z\in\chi_\emptyset} z \ \mb{S}^{(+)}(z)\prod_f \mb{V}(\bar Q_f)\ket{\mu_0}}=\\
=\sum_{\vec Y}\ Q_g^{|\vec Y|}\ \frac{\prod_{A,f}N_{\emptyset Y_A }(t^{-1/2}p^{1/2}\bar Q_f/x_A;q,t)N_{Y_A\emptyset }(q^{1/2} x_A /Q_f,;q,t)}{\prod_{A,B} N_{Y_A Y_B}(x_{AB};q,t)}\ .
\end{multline}

The standard relation between vortex and instanton partition functions (see \textit{e.g.} \cite{Aganagic:2013tta}) allows one to identify the two approaches at \textit{specific limits} of the 5d Coulomb branch parameters. In fact, at $q^{1/2} x_A/Q_f=t^{r_A}$,  $r_A\in\mathbb{Z}_{\geq 0}$, only Young diagrams $Y_A$ with at most $r_A$ rows contribute to the instanton partition function, and  (\ref{VZVinst}) collapses to the vortex partition function (\ref{HZvortex}) with $r=\sum_A r_A$ and  normalized by its perturbative part. We refer to \cite{Aganagic:2013tta} for more details about the identification. 

\subsubsection*{Relation between contour and Jackson integrals}
We would like to close this section by briefly discussing a formal relation between ordinary contour integrals and Jackson integrals. This relation will produce a map between the screening charges adopted here and those in section \ref{sec:qVir}. 

The ordinary definite Jackson integrals are defined by
\be
\begin{split}
\int_0^z\d_q x\; f(x)&\equiv(1-q)\sum_{k\geq 0}z q^k f(z q^k)~,\\
\int_{z}^\infty\d_q x\; f(x)&\equiv \int_0^{z^{-1}}\frac{\d_q y}{y^2}\; f(y^{-1})=(1-q)\sum_{k\geq 0}z q^{-k}f(z q^{-k})~.
\end{split}
\ee
We define the {\it based Jackson integral}  to be (without  the $1-q$ factor for simplicity)
\be
\int_{z}\d_q x\; f(x)\equiv\frac{1}{1-q}\left(\int_0^z \d_q x\; f(x)+\int_{z q^{-1}}^\infty \d_q x\; f(x)\right)=\sum_{k\in\mathbb{Z}} \ aq^k  f(z q^k)~.
\ee
Notice that when $z=1$, this definition coincides with the improper Jackson integral
\be
\int_{1} \d_q x\; f(x)=\frac{1}{1-q}\int_0^\infty\d_q x\; f(x)=\sum_{k\in\mathbb{Z}}q^k f(q^k)~.
\ee
We can give a relation between based Jackson integrals and ordinary contour integrals by using $q$-constants. For instance, let us consider the $q$-constant
\be
c_\lambda(x;q)=x^\lambda\frac{\Theta(q^\lambda x;q)}{\Theta(x;q)} \ ,\quad c_\lambda(q x;q)=c_\lambda(x;q) \ , \quad \lambda\in\mathbb{C}\backslash\mathbb{Z}~.
\ee
If we assume the function $f(x)$ to be regular at $x=z q^{\mathbb{Z}}$, then we have
\be
\oint\frac{\d x}{2\pi\i} \ c_\lambda(x /z;q)\; f(x)={\rm Res}_{x=1}\frac{\Theta(q^\lambda;q)}{\Theta(x;q)}\sum_{k\in\mathbb{Z}}z q^k f(z q^k)~,
\ee
where the integration contour is chosen to pick up the sum of the residues at the poles $x=z q^\mathbb{Z}$ coming from the zeros of the denominator of $c_\lambda(x / z;q)$. Assuming $|q|<1$, this means we are integrating around a segment interpolating between $x=0$ and $x=\infty$ passing through $x=z$. In fact, for $k\geq 0$ we integrate around the segment $[0,z]$, while for $k<0$ we integrate around the segment $[q^{-1} z,\infty)$. This fits with our definition of the based Jackson integral, which is then given by
\be
\int_{z}\d_q x\ f(x)=-\frac{(q;q)^2_\infty}{\Theta(q^\lambda;q)}\oint\frac{\d x}{2\pi\i }c_\lambda(x/ z;q)\; f(x)~,
\ee
where we used ${\rm Res}_{x=1}\Theta(x;q)^{-1}=-(q;q)^{-2}_\infty$. If we extend the based Jackson integral to operator-valued functions, we can write the screening charge (\ref{Qjack}) as 
\be
\mb{Q}_{z} = \int_{z}\d_q x\ \mb{S}(x)=-\frac{(q;q)^2_\infty}{\Theta(q^\lambda;q)}\oint\frac{\d x}{2\pi\i }\ c_\lambda(x /z;q)\ \mb{S}(x)~.
\ee
Moreover,  $\lambda$ has appeared so far as a free parameter, and then we can try to turn it into the  $\mb{P}$ operator\footnote{Also,  we are always evaluating free boson correlators in a basis diagonalizing $\mb{P}$.} and consider 
\be\label{QSP}
\mb{Q}_z\equiv -(q;q)^2_\infty\oint\frac{\d x}{2\pi\i }\ \mb{S}(x) \frac{c_{-\sqrt{\beta}\mb{P}}(x /z;q)}{{\Theta(q^{-\sqrt{\beta}\mb{P}};q)}} \ .
\ee
Notice that the zero mode part of the integrand is
\be
\Bigg[\mb{S}(x)\frac{c_{-\sqrt{\beta}\mb{P}}(x/ z;q)}{\Theta(q^{-\sqrt{\beta}\mb{P}};q)}\Bigg]_0=\e^{\sqrt{\beta}\mb{Q}}\ \frac{\Theta(q^{-\sqrt{\beta}\mb{P}}x / z;q)}{\Theta(x /z;q)\Theta(q^{-\sqrt{\beta}\mb{P}};q)} \ z^{\sqrt{\beta}\mb{P}}~,
\ee
which, at $z=1$, is exactly the redefinition of the zero mode part that was introduced in \cite{Lodin:2017lrc} (see also \cite{Nieri:2017vrb} for more explanations). We also observe that the integrand appearing in (\ref{QSP}) is equivalent to a \textit{dressed} screening current given by
\be
 \mb{S}(x)\, c_{-\sqrt{\beta}\mb{P}}(x /z;q)=\mb{\Phi}_{\mb{P}-2\sqrt{\beta}}(z)^{-1}\mb{S}(x)\mb{\Phi}_{\mb{P}}(z) \ ,\quad 
\mb{\Phi}_{\mb{P}}(z)\equiv \ :\mb{V}(q^{1/2}z q^{\sqrt{\beta}\mb{P}}) \mb{V}(q^{1/2}z)^{-1}:  .
\ee

%%%%%%%%%%  Bibliography  %%%%%%%%%%%%

\bibliographystyle{utphys}
%\bibliography{ref}

\providecommand{\href}[2]{#2}\begingroup\raggedright\endgroup

\end{document}